\documentclass[11pt]{article}
\pdfoutput=1
\usepackage{graphicx,amssymb,amsfonts,amsmath,amssymb,amscd,amstext, mathrsfs}
\usepackage{color}
\makeatletter

\@addtoreset{equation}{section}
\newcommand{\be}{\begin{eqnarray}}
\newcommand{\ee}{\end{eqnarray}}
\newcommand{\ba}{\begin{array}}
\newcommand{\ea}{\end{array}}
\newcommand{\nn}{\nonumber}

\renewcommand{\(}{\Big(}
\renewcommand{\)}{\Big)}

\def \<{\langle}
\def \>{\rangle}

\makeatletter \@addtoreset{equation}{section} \makeatother

\begin{document}

\vspace{0.6cm}

\begin{center}
~\\~\\~\\
{\bf  \LARGE Perturbative Complexity of Interacting Theory}
\vspace{2cm}

Wung-Hong Huang*\\
\vspace{0.5cm}
Department of Physics, National Cheng Kung University,\\
No.1, University Road, Tainan 701, Taiwan

\end{center}
\vspace{2cm}
\begin{center}{\bf  \large Abstract}\end{center}
We present a systematic method to expand the quantum complexity of interacting theory in  series of coupling constant.  The complexity is evaluated by the operator approach in which the transformation matrix between the  second quantization operators of reference state and the target state defines the quantum gate.  We start with  two coupled oscillators and  perturbatively evaluate the geodesic length of the associated group manifold of  gate matrix. Next, we generalize the analysis to $N$ coupled oscillators which describes the lattice $\lambda\phi^4$ theory.  Especially, we introduce  simple diagrams  to represent the perturbative series and construct simple rules to efficiently calculate the complexity.  General formulae are obtained for the higher-order complexity of excited states.  We present several diagrams to  illuminate the properties of complexity and show that the interaction correction  to complexity may be positive or negative depending on the magnitude of reference-state frequency.

\vspace{4cm}
\begin{flushleft}
* Retired Professor of NCKU, Taiwan. \\
* E-mail: whhwung@mail.ncku.edu.tw
\end{flushleft}
\newpage
\tableofcontents
\section{Introduction}

Achieving a better understanding of physics behind a black-hole horizon is important if one wants to precisely describe the bulk geometry in terms of the information of boundary CFT  \cite{Raamsdonk1005, Swingle1405, Lashkari1308, Faulkner1312, Maldacena0106, Hartman1305, Maldacena1306}.  
 In the context of the eternal AdS-Schwarzchild black hole, for instance, a related question concerns the size of a wormhole growing linearly with time; this phenomenon has been conjectured to dual to the growth of ``complexity" of the dual CFT \cite{Susskind1403}.   In the complexity=volume (CV) conjecture \cite{Susskind1403},  the  complexity is dual to the volume of an extremal codimension-one bulk surface anchored to a  time slice of the boundary. In  the complexity=action (CA) conjecture  \cite{Carmi1709, Brown1509, Brown1512, Chapman1701}, one identifies the complexity with a gravitational action evaluated on the Wheeler-DeWitt (WDW) patch, anchored also on a time slice of the boundary. 

Several efforts were made to provide a definition of the complexity in the field theory \cite{Jefferson1707, Chapman1707, Khan1801, Hackl1803, Bhattacharyya1880}.  The complexity in there is defined as the number of  operations $\{\cal O^I\}$ needed to transform a reference state $|\psi_R\rangle$ to a target state $|\psi_T\rangle$.   These operators are also called as quantum gates: the more gates one needs, the more complex the target state is.  
One can define an  affine parameter ``s" associated to an unitary operator $U(s)$ and  use a set of function, $Y^I(s)$, to character the quantum circuit. 
The unitary operation connecting the reference state and target state is 
\be
U(s)=\vec{\cal P}\, e^{\int_0^s\,Y_I(s)\,{\cal O}_I},~~~~|\psi_R\rangle =U(0)|\psi_R\rangle,~~~~~|\psi_T\rangle =U(1)|\psi_R\rangle~~~\label{gate} \ ,
\ee
where $\vec{\cal P}$ is a time ordering along $s$. The  complexity ${\cal C}$ and circuit depth $D[U]$ (cost function) are  \cite{Jefferson1707} 
\be
{\cal C}&=&\underset{\rm \{Y^I\}}{\rm Min}\,D[U],~~~~~~D[U]=\int_0^1ds\,{\sum_I |Y^I(s)|^2} \ .~~~\label{distance}
\ee 
Above definitions were shown to  be consistent with a gravitational computation  \cite{Jefferson1707}.  
The initial studies in field theory considered the Gaussian ground state wavefunctions in reference state and target state \cite{Jefferson1707,Khan1801, Hackl1803}.  The theories studied so far are  the free field theory  or exponential type  wavefunction in interacting model \cite{Bhattacharyya1880}. The operator approach had also been used in \cite{Chapman1707, Hackl1803} to study the complexity of  fermion theory. 

In our previous paper  \cite{Huang1905} we adopt the operator approach, in which the transformation matrix between the second quantization operators of reference state and target state is regarded as the quantum gate, to evaluate the complexity in free scalar field theory.   Since that in the  operator approach we need not to use the explicit form of the wave function we can study the complexity in the excited states\footnote{Note that the excited-state wavefunction of harmonic oscillation  is not pure  exponential form and  the wavefunction approach is hard to work.}.  We first  examined the system in which the reference state is  two oscillators with same frequency $\omega_f$ while the  target state  is  two  oscillators with frequency $ \omega_1$ and $ \omega_2$.  We explicitly calculated the complexity in several excited states  and  proved that the square of geodesic length in the general state $|{\rm N_1,N_2}\rangle$ is
\be
D_{\rm (N_1,N_2)}^2={\rm (N_1+1)}\left(\ln {\sqrt{\omega_1\over  \omega_f}}\,\right)^2 +{\rm (N_2+1)}\left(\ln {\sqrt{ \omega_2\over  \omega_f}}\,\right)^2
\ee
The results was furthermore  extended to the N couple harmonic oscillators which correspond to  the lattice version of  free scalar field,  see  sec.5 of \cite{Huang1905}. 

In this paper we extend \cite{Huang1905} by including interactions to further study the complexity using the operator approach.  We present a systematic method to evaluate the complexity of the $\lambda\phi^4$ field theory by the perturbation of small coupling constant.   An outline of the paper is as follows.

 In section II,  as that in \cite{Jefferson1707}  we describes the lattice  scalar field as coupled oscillators.  In section III we consider two coupled oscillators  and  find that, to the $\lambda^n$ order  the square distance of excited state between target and reference state is
\be
D^{(n)2}_{(N_1,N_2)}=(N_1+1)\left(\ln \(\sqrt {R^{(n)}_1}\)\right)^2+(N_2+1)\left(\ln \(\sqrt {R^{(n)}_2}\)\right)^2
\ee
in which $R^{(n)}_1$ and $R^{(n)}_2$ are described in (\ref{R2}).  In section IV we  generalize it to the case of N coupled oscillators which correspond to  the lattice version of  $\lambda\phi^4$ theory.   We use  new kind of simple diagrams, figures 3, 4 and 5,  to represent the perturbative series and construct  simple rules, figures 1 and 2, to calculate the complexity therein.  We find that the diagrams are classified into  three classes : odd $N$, odd ${N\over 2}$, and even ${N\over 2}$. We explicitly calculate the complexity in the cases of N=2,3,4, 5 to any order of $\lambda$.   Using these experiences we then in section V  derive the general formulas of complexity in (\ref{F1}), (\ref{F2}), and (\ref{F3}).  Then, we present several  diagrams to  illuminate the properties of complexity and find that the interaction correction  to complexity may be positive or negative depending on the magnitude of reference-state frequency.  We conclude in Sec. 6. 

\section{Interacting  Scalar Field and Coupled Oscillators}

The $d$-dimensional massive scalar Hamiltonian with a $\hat\lambda \phi^4$ interaction is
\be
H=\frac{1}{2}\int d^{d-1}x\left[\pi(x)^2+\vec\nabla\phi(x)^2+m^2\phi(x)^2+{\hat\lambda\over 12}\phi(x)^4\right] \ .
\ee
Placing the theory on a square  lattice with lattice spacing  $\delta$,  one has
\be
H=\frac{1}{2}\sum_{\vec n}\left\{\frac{p(\vec n)^2}{\delta^{d-1}}+\delta^{d-1}\Big[\frac{1}{\delta^2}\sum_i\Big((\phi(\vec n)-\phi(\vec n-\hat{a}_i))^2+m^2\phi(\vec n)^2+{\hat\lambda\over 12}\phi(\vec n)^4\Big]\right\} \ ,
\ee
where $\hat{a}_i$ are unit vectors pointing toward the spatial directions of the lattice. By redefining 
\be
&& X(\vec n)=\delta^{d/2}\phi(\vec n), ~~~ P(\vec n)={p(\vec n)\over \delta^{d/2}}, ~~ M={1\over \delta}, ~~ \omega=m, ~~~ \Omega={1\over \delta}, ~~~ \lambda={\hat \lambda \over 24\delta^4} \ , 
\ee 
the Hamiltonian becomes
\be
H=\sum_{\vec n}\left\{\frac{P(\vec n)^2}{2M}+\frac12 M \left[\omega^2 X(\vec n)^2+\Omega^2\sum_i\Big( X(\vec n)-X(\vec n-\hat{a}_i)\Big)^2+2\lambda\,X(\vec n)^4\right]\right\} \ ,
\ee 
When $\vec n$ is an one dimensional vector the Hamiltonian describes an infinite family of coupled one dimensional  oscillators.  We will extensively study the one dimensional  oscillators in this paper while the extension to higher dim is just to replace the site index ``$\,\,i\,\,$''  to  ``$\,\,\vec i\,\,$'', as that described in  \cite{Jefferson1707}.
\section{Two Coupled Oscillators}
First we consider a simple case of two coupled  oscillators ($M=1$):
\be
H&=&{1\over 2}\Big[\tilde p_1^2+\tilde p_2^2+\omega^2(\tilde x_1^2+\tilde x_2^2)+\Omega^2(\tilde x_1-\tilde x_2)^2+2\lambda(\tilde x_1^4+\tilde x_2^4)\Big]
\ee
Defining
\be
\tilde x_{1,2}&=&{1\over \sqrt 2}(x_1\pm x_2),~~\tilde p_{1,2}={1\over \sqrt 2}(p_1\pm p_2),~~~ \omega_1^2=\omega^2, ~~~ \omega_2^2=\omega^2+2\Omega^2 \ ,
\ee
the Hamiltonian is
\be
H &=&{1\over 2}\( p_1^2+ \omega_1^2 x_1^2+ p_2^2+ \omega_2^2 x_2^2\)+{\lambda\over 4}\,\((x_1+ x_2)^4+( x_1- x_2)^4\)=K+V
\ee
In the second quantization, we define
\be
 a_1^\dag&=&\sqrt{\omega_1\over 2}\, x_1+i{1\over \sqrt{2\omega_1}} p_1,~~~~ a_2^\dag =\sqrt{\omega_2\over 2}\,x_2+i{1\over \sqrt{2\omega_2}} p_2,~~~~[a_{1,2},\,a^\dag_{1,2}]=1\\
 x_{1,2}&=&\sqrt{1\over 2\omega_{1,2}}( a_{1,2}^\dag+ a_{1,2}),~~~~ p_{1,2}=i\sqrt{{\omega_{1,2}\over  2}}( a_{1,2}^\dag- a_{1,2}) \ .
\ee
The state wavefunction is $\psi(x_{1}, x_{2})=\<x_1,x_2| a_1^\dag a_2^\dag\,|0\>$.  
\subsection{Kinetic Term of Two Coupled Oscillators}

The  kinetic term has a diagonal form:
\be
K^{\rm (tar)}&=&\omega_1 a_1^\dag  a_1+\omega_2 a_2^\dag  a_2+{1\over 2} (\omega_1+\omega_2)
\ee
where the constant terms are irrelevant to the following discussions.  
We choose the reference state with the associated kinetic term given by \cite{Jefferson1707}
\be
K^{\rm (ref)}&=&\omega_f\,(a^{\rm (ref)}_1)^\dag  a^{\rm (ref)}_1+\omega_f\,(a^{\rm (ref)}_2)^\dag  a^{\rm (ref)}_2 \ .
\ee
Now we see that with the replacement
\be
a^{\rm (ref)}_{1}\rightarrow \sqrt {\omega_{1}\over \omega_f}\,a_{1},~~a^{\rm (ref)}_{2}\rightarrow \sqrt {\omega_{2}\over \omega_f}\,a_{2}~~~\label{rep}
\ee
one can obtain $K^{\rm (tar)}$ from  $K^{\rm (ref)}$, i.e.
\be
 K^{\rm (\rm ref)}\rightarrow  K^{\rm (tar)} 
\ee
In the operator approach the gate matrice defined in (\ref{gate}) is constructed by the transformation from target operator to referenct operator in above relation. 

Consider first the ground state which is annihilated by  $a_1,a_2$, i.e. $a_1a_2|0,0\>=0$ for target state, and  $a^{\rm (ref)}_1a^{\rm (ref)}_2|0,0\>_{\rm ref}=0$ for reference state. The matrix $U(s)$ connecting the target operator with  referenct operator  in  (\ref{gate}) is
\be
U(1)=\left(
\ba{cc}
\sqrt {\omega_{1}\over \omega_f}&0\\
0&\sqrt {\omega_{2}\over \omega_f}\\
\ea\right)
\Rightarrow~
\left(
\ba{cc}
a_1&a_2\\
\ea\right)
=
U(1)
\left(
\ba{c}
a^{\rm (ref)}_1\\
a^{\rm (ref)}_2\\
\ea\right)~~~\label{U}
\ee
which lead to the replacement relation in (\ref{rep}), with initial condition $U(0)=diag(1,1)$.  Since the  transformation matrix $U(1)$ is diagonal we can choose  ${\cal O_I}=1$ in (\ref{gate}) and  have a simple relation
\be
U(1)=\left(
\ba{cc}
e^{\int_0^1\,ds \,Y_1(s)}&0\\
0&e^{\int_0^1\,ds \,Y_2(s)}\\
\ea\right)
=\left(
\ba{cc}
\sqrt {\omega_{1}\over \omega_f}&0\\
0&\sqrt {\omega_{2}\over \omega_f}\\
\ea\right)
\ee
The associated solutions of $Y_{(1,2)}(s)$  are
\be
Y_1(s)&=&\ln\(\sqrt {\omega_1\over \omega_f}\,\),~~~Y_2(s)=\ln\(\sqrt {\omega_2\over \omega_f}\,\)
\ee
which satisfied the initial condition.   The squared distance for ground state, denoted as $D^2_{(0,0)}$, between target and reference state calculated from  (\ref{distance}) becomes
\be
D^2_{(0,0)}=Y_1(1)^2+Y_2(1)^2=\left(\ln \(\sqrt {\omega_1\over \omega_f}\)\right)^2+\left(\ln \(\sqrt {\omega_2\over \omega_f}\)\right)^2~~~~\label{d0}
\ee
This matches with the result obtained earlier in \cite{Huang1905}.

Consider next the $\{N_1^{th}\,N_2^{th}\}$ excited state which is defined by $a^{N_1+1}_1a^{N_2+1}_2|N_1,N_2\>=0$, or $|N_1,N_2\>={(a^\dag_1)^{N_1+1}(a^\dag_2)^{N_2+1}\over \sqrt{(N_1+1)!\,(N_2+1)!}}|0,0\>$. In this case the gate matrices can be  read from the transformations 
\be
\overbrace{a^{\rm (ref)}_{1}\,....a^{\rm (ref)}_{1}}^{N_1+1}\,\overbrace{a^{\rm (ref)}_{2}\,....a^{\rm (ref)}_{2}}^{N_2+1}\,\,\Rightarrow\,\,\overbrace{\sqrt {\omega_{1}\over \omega_f}\,a_{1}\,....\sqrt {\omega_{1}\over \omega_f}\,a_{1}}^{N_1+1}\,\overbrace{\sqrt {\omega_{2}\over \omega_f}\,a_{2}\,....\sqrt {\omega_{2}\over \omega_f}\,a_{2}}^{N_2+1}\,
\ee
Then, the operator  connecting the target operator with  referenct operator  in  (\ref{gate}) becomes a $(N_1+1)\times (N_2+1)$ diagonal matrix $U(s)$ with elements
\be
U(1)=diag\left(\overbrace{\sqrt {\omega_1\over \omega_f},...,\sqrt {\omega_1\over \omega_f}}^{N_1+1}\,,\overbrace{\sqrt {\omega_2\over \omega_f},...,\sqrt {\omega_2\over \omega_f}}^{N_2+1}\right)
\ee
which becomes (\ref{U}) in the case of ground state $N_1=N_2=0$. Follow the discussions in before the  matrix $U_i$ defined in  (\ref{gate}) now becomes
\be
U_i(1)&=& \sqrt {\omega_{1}\over \omega_f}\,,~~{\rm with}~~U_{i}(0)=1,~~1\le i\le N_1+1\\
U_j(1)&=&\sqrt {\omega_{2}\over \omega_f}\,,~~{\rm with}~~U_{j}(0)=1,~~~~N_1+2\le j\le N_2+N_1+2
\ee
The associated functions of $Y_i(s)$ solved from   (\ref{gate}) are
\be
Y_i(s)&=&\ln\( \sqrt {\omega_1\over \omega_f}\,\),~~1\le i\le N_1+1\\
Y_j(s)&=&\ln\( \sqrt {\omega_2\over \omega_f}\,\),~~,~~~~N_1+2\le j\le N_2+N_1+2
\ee
The squared distance for excited  state, denoted as $D^2_{(N_1,N_2)}$, between target and reference state calculated from  (\ref{distance}) is
\be
D^2_{(N_1,N_2)}&=&\sum_{i=1}^{N_1+1}Y_i(1)^2+\sum_{j=N_1+2}^{N_1+N_2+2}Y_j(1)^2\nn\\
&=&(N_1+1)\left(\ln \(\sqrt {\omega_1\over \omega_f}\)\right)^2+(N_2+1)\left(\ln \(\sqrt {\omega_2\over \omega_f}\)\right)^2 \ .
\ee
This matches with the result obtained earlier in \cite{Huang1905}. 

 Recall that the state wavefunction is described by $\Psi_n(x)={1\over \sqrt{n!}}\<x|(a^\dag)^n|0\>$  the gate matrix of  excited-state wavefunction, $\Psi_n(x)$, is thus related to the gate matrix of  field operators, $(a^\dag)^n$.

\subsection{Interacting Term of Two Coupled Oscillators}
We next study the correction to the complexity due to the interaction term:
\be
V&=&{\lambda\over 4}\,\((x_1+ x_2)^4+( x_1- x_2)^4\)\\
&=&{\lambda\over 4\cdot 2^2}\left[\(\sqrt{1\over \omega_{1}}( a_{1}^\dag+ a_{1})+\sqrt{1\over \omega_{2}}( a_{2}^\dag+ a_{2})\)^4\(\sqrt{1\over \omega_{1}}( a_{1}^\dag+ a_{1})-\sqrt{1\over \omega_{2}}( a_{2}^\dag+ a_{2})\)^4\right]\nn\\
&=&{\lambda\cdot2\over 4\cdot 2^2}\left[\(\sqrt{1\over \omega_{1}}( a_{1}^\dag+ a_{1})\)^4+\(\sqrt{1\over \omega_{2}}( a_{2}^\dag+ a_{2})\)^4+6\(\sqrt{1\over \omega_{1}}( a_{1}^\dag+ a_{1})\)^2\(\sqrt{1\over \omega_{2}}( a_{2}^\dag+ a_{2})\)^2\right]\nn
\ee
We will consider $\<N_1,N_2|V|N_1,N_2\>$ for the excited state $|N_1,N_2\>$ with fixed $N_1$ and $N_2$. 
In this way, only the terms that have the same power of $a_i$ and $a^\dag_i$ are relevant.  Therefore we only need to consider
\be
\(a_{1}^\dag+ a_{1}\)^4&=&\((a_1^\dag)^2+a_1^\dag a_1+a_1a_1^\dag+(a_1)^2\)\((a_1^\dag)^2+a_1^\dag a_1+a_1a_1^\dag+(a_1)^2\)\nn\\
&=&6a_1^\dag a_1a_1^\dag a_1+6a_1^\dag a_1+3+{\rm irrelevant~terms}~~\label{a4}\\
\(a_{1}^\dag+ a_{1}\)^2&=&(a_1^\dag)^2+a_1^\dag a_1+a_1a_1^\dag+(a_1)^2
= 2a_1^\dag a_1+1+{\rm irrelevant~terms}~~\label{a2} \ .
\ee
We obtain, after dropping irrelevant terms,
\be
H
=\left(\omega_1+{3\lambda\over 2}\,\({1+N_1\over 2\omega^2_1}+{1+N_2\over \omega_1\omega_2}\)\right) a_1^\dag  a_1+
\left(\omega_2+{3\lambda\over 2}\({1+N_1\over \omega_2\omega_1}+{1+N_2\over 2\omega^2_2}\)\right) a_2^\dag  a_2 \ .~~\label{VN=2}
\ee
The associated  Hamiltonian of the reference state can be chosen as   
\be 
H^{\rm (ref)}&=&\left(\omega_f+{3\lambda\over 2}\,{1+N_1\over 2\omega^2_f}\right) (a_1^{\rm (ref)})^\dag  a^{\rm (ref)}_1+\left(\omega_f+{3\lambda\over 2}\,{1+N_2\over 2\omega^2_f}\right) (a_2^{\rm (ref)})^\dag  a^{\rm (ref)}_2 
\ee
which satisfies a desirable property of the reference state that it does not contain any entanglement between 
operators $\{N_1,\, a_1^{\rm (ref)},\,  (a_1^{\rm (ref)})^\dag\} $ and $\{N_2,\, a_2^{\rm (ref)},\,  (a_2^{\rm (ref)})^\dag\} $. The property is like as that in coordinate approach in which a desirable property of the reference state is that it should not contain any entanglement between the original coordinates $x_1$ and $x_2$ \cite {Bhattacharyya1880}.

In the case of zero-order of $\lambda$,
\be
K^{\rm (ref)}=\omega_f\,(a_1^{\rm (ref)})^\dag  a^{\rm (ref)}_1+\omega_f\, (a_2^{\rm (ref)})^\dag  a^{\rm (ref)}_2,~~~K^{\rm (tar)}=\omega_1\, a_1^\dag\,a_1+\omega_1\, a_2^\dag\,a_2
\ee
which implies transformations
\be
a^{\rm (ref)}_{1}\to \sqrt{\omega_{1}\over \omega_f}\,a^{\rm (ref)}_{1},~~a^{\rm (ref)}_{2}\to \sqrt{\omega_{2}\over \omega_f}\,a^{\rm (ref)}_{2} &{\Longrightarrow}&K^{\rm (ref)} \to  K^{\rm (tar)}\\
{\rm or}~~~ N^{\rm (ref)}_{(1,2)}\to R^{(0)}_{(1,2)}\,N^{\rm (ref)}_{(1,2)} &{\Longrightarrow}& K^{\rm (ref)} \to  K^{\rm (tar)}~~\label{N}
\ee
where
\be
R^{(0)}_i&=&{\omega_i\over \omega_f}~~\label{R0}~~~~~~\label{ratio}
\ee
The quantum gate are described by two $1\times 1$ matrices,  $ {\rm exp}\(\sqrt{R^{(0)}_1}\)$ and $ {\rm exp}\(\sqrt{R^{(0)}_2}\)$.  
This is the case of purely kinetic term, i.e. a free theory. 

Now consider a perturbation to the complexity for the two coupled oscillators.   At the first order of $\lambda$,  we have transformations
\be
\left\{\ba{ccc}
\left(\omega_1+{3\lambda\over 2}\,\({1+N_1\over 2\omega^2_1}+{1+N_2\over \omega_1\omega_2}\)\right) a_1^\dag  a_1&\rightarrow&\left(\omega_f+{3\lambda\over 2}\,{1+N_1\over 2\omega^2_f}\right) (a_1^{\rm (ref)})^\dag  a^{\rm (ref)}_1\\
\\
\left(\omega_2+{3\lambda\over 2}\({1+N_1\over \omega_2\omega_1}+{1+N_2\over 2\omega^2_2}\)\right) a_2^\dag  a_2&\rightarrow&\left(\omega_f+{3\lambda\over 2}\,{1+N_2\over 2\omega^2_f}\right) (a_2^{\rm (ref)})^\dag  a^{\rm (ref)}_2 \ . 
\ea  \right.    
\ee
The factors $N_{(1,2)}$ are within the coupling term, i.e. ${3\lambda\over 2}$, and we only need to consider their zero-order transform. Recall (\ref{N}), we have to multiple them by $R^{(0)}_{(1,2)}$ factors.  Therefore the first-order transformations are
\be
R^{(1)}_1&=&{\omega_1+{3\lambda\over 2}\,\({1+N_1R^{(0)}_1\over 2\omega^2_1}+{1+N_2R^{(0)}_2\over \omega_1\omega_2}\)\over  \omega_f+{3\lambda\over 2}\,{1+N_1\over 2\omega^2_f}},~~~~
R^{(1)}_2={\omega_2+{3\lambda\over 2}\({1+N_1R^{(0)}_1\over \omega_2\omega_1}+{1+N_2R^{(0)}_2\over 2\omega^2_2}\)
\over   \omega_f+{3\lambda\over 2}\,{1+N_2\over 2\omega^2_f} }~~~\label{R21} \ ,
\ee
Now, the functions $(R^{(1)}_1,\,R^{(1)}_2)$ play the roles of $({\omega_1\over \omega_f},\,{\omega_2\over \omega_f})$  in free theory, i.e. (\ref{ratio}), and the square distance formula (\ref{d0}) becomes
\be
D^{(1)2}_{(0,0)}=\left(\ln \(\sqrt {R^{(1)}_1}\)\right)^2+\left(\ln \(\sqrt {R^{(1)}_2}\)\right)^2 \ .
\ee
 For excited states,  along the same analysis in free theory, the first-order square distance is
\be
D^2_{(N_1,N_2)}=(N_1+1)\left(\ln \(\sqrt {R^{(1)}_1}\)\right)^2+(N_2+1)\left(\ln \(\sqrt {R^{(1)}_2}\)\right)^2 \ .
\ee
Extending to higher-order interactions is straightforward.  The recursion relations are
\be
R^{(n)}_1={\omega_1+{3\lambda\over 2}\,\({1+N_1R^{(n-1)}_1\over 2\omega^2_1}+{1+N_2R^{(n-1)}_2\over \omega_1\omega_2}\)\over  \omega_f+{3\lambda\over 2}\,{1+N_1\over 2\omega^2_f}},~~
R^{(n)}_2={\omega_2+{3\lambda\over 2}\({1+N_1R^{(n-1)}_1\over \omega_2\omega_1}+{1+N_2R^{(n-1)}_2\over 2\omega^2_2}\)
\over   \omega_f+{3\lambda\over 2}\,{1+N_2\over 2\omega^2_f}}  \ ~~~\label{R2}
\ee
with initial values  $R^{(0)}_{(1,2)}$ defined in (\ref{R0}).  
For excited states, the $n$-order square distance is
\be
D^{(n)2}_{(N_1,N_2)}=(N_1+1)\left(\ln \(\sqrt {R^{(n)}_1}\)\right)^2+(N_2+1)\left(\ln \(\sqrt {R^{(n)}_2}\)\right)^2 \ ,
\ee 
which is the $n$-order complexity of two coupled oscillators.

Note that  original relations (\ref{R2}) can  be expanded  as 
\be
R^{(n)}_1&\approx&{\omega_1\over \omega_f}+{3\lambda\over 2 \omega_f}\,\({1+N_1R^{(n-1)}_1\over 2\omega^2_1}+{1+N_2R^{(n-1)}_2\over \omega_1\omega_2}-{1+N_1\over 2\omega^2_f}\)~~\label{R2s1}\\
R^{(n)}_2&\approx&{\omega_2\over \omega_f}+{3\lambda\over 2 \omega_f}\,\({1+N_1R^{(n-1)}_1\over \omega_2\omega_1}+{1+N_2R^{(n-1)}_2\over 2\omega_2^2}-{1+N_2\over 2\omega^2_f}\)~~\label{R2s2}
\ee
In this way, the perturbative series of $R^{(n)}_i$ in coupling constant $\lambda$ is explicitly showing up.  However, to save the space,  in following sections we will not expand the original relations, likes  as (\ref{R2}), to the relations, likes  as (\ref{R2s1}) or (\ref{R2s2}).
\section{N Coupled Oscillators}
\subsection{Kinetic Term of N Coupled Oscillators}
For $N$ coupled  oscillators, 
\be
H&=&{1\over 2}\sum_{k=1}^{N}\Big[\tilde p_k^2+\omega^2\tilde x_k^2+\Omega^2(\tilde x_k-\tilde x_{k+1})^2+2\lambda\,\tilde x_k^4\Big] \ .
\ee
We impose a periodic boundary condition $\tilde x_{\rm k+N+1}=\tilde x_k$. The normal coordinates are chosen to be 
\be
x_k&=& {1\over \sqrt N}\sum_{j=1}^{N}\,exp\Big({2\pi i k\over N} \,j\Big)\,\tilde x_j,~~~~ p_k={1\over \sqrt N}\sum_{j=1}^{N}\,exp\Big({-2\pi i k\over N} \,j\Big)\,\tilde p_j 
\ee
Note that the  relative sign between the Fourier series of $x_k$ and $p_k$ is important to have standard commuation relation $[x_{k_1},p_{k_2}]=\delta_{k_1,k_2}$ \cite{Jefferson1707}.    
The Hamiltonian now becomes
\be
H={1\over2}\sum_{k=1}^{N}\, \(p_k^\dag p_k +\omega_k^2\,x_k^\dag x_k\)+V, ~~~~~ \omega_k^2= \omega^2+4\Omega^2\,\sin^2{\pi k\over N}~~~~\label{frequency}
\ee
Defining
\be
x_k={1\over \sqrt{2\omega_k}}(a_k+a^\dag_{-k}),~~~p_k=i\sqrt{\omega_k\over 2} (a^\dag_k-a_{-k}),~~~~[a_{k},\,a^\dag_{k}]=1 \ ,
\ee
the kinetic term can be written as
\be
\sum_kK_k= {1\over 2}\sum_k\(p_k^\dag p_k+\omega_k^2\,x_k^\dag x_k\)=\sum_k{\omega_k}\,a^\dag_k a_k~~\label{K}\ , 
\ee 
up to an irrelevant constant.

The states in $N$ oscillators can be defined by the creation operators $ a_{1}^\dag  a_{2}^\dag... a_{k}^\dag...$,such that  $\psi(x_{1}, x_{2},...)=\<x_1,x_2,..x_k....|  a_{1}^\dag  a_{2}^\dag... a_{k}^\dag...|0\>$. 
As before, to find the complexity of such state we choose a reference state with the associated kinetic term given by
\be
K^{\rm (ref)}&=&\sum_k \omega_f\,(a^{\rm (ref)}_k)^\dag  a^{\rm (ref)}_k \ .
\ee
The square distance for the  $n_k$-th excited state is 
\be
D^2_{\{N_1,N_2,..N_k,..N_N\}}&=&(N_1+1)\left(\ln \(\sqrt {\omega_{1}\over \omega_f}\)\right)^2+(N_2+1)\left(\ln \(\sqrt {\omega_{2}\over \omega_f}\)\right)^2+ \dots\nn\\
&=&\sum_{k=1}^{N}\,(N_k+1)\left(\ln \(\sqrt {\omega_{k}\over \omega_f}\)\right)^2
\ee
where $\omega_k$ is defined in (\ref{frequency}).
\subsection{Interacting  Term of N Coupled Oscillators : Perturbative Algorithm}

We adopt the following steps to systematically study a perturbation theory of the complexity:
\\

 (I)  We express potential $V$ in terms of $a, \, a^\dag$: 
\be
V&=&\lambda\sum_{k=1}^{N}\,\tilde x_k^4=\lambda\sum_{k=1}^{N}\,\({1\over \sqrt N}\sum_{j=1}^{N}\, \exp\Big({-2\pi i k\over N} \,j\Big)\, x_j\)^4\nn\\
&=&{\lambda\over 4N^2}\sum_{k=1}^{N}\,\left(\sum_{j=1}^{N}\, \exp\Big({-2\pi i k\over N} \,j\Big)\, \({1\over \sqrt{\omega_j}}(a_j+a^\dag_{j})\)\right)^4~~\label{potential}
\ee

(II) Define
\be
A(j)={1\over \sqrt{\omega_j}}(a_j+a^\dag_{j}) \ .
\ee
Then,  as calculated in (\ref{a4}) and (\ref{a2}), 
\be
A(j)^4&=&{6\over \omega_j^2}(a_j^\dag a_ja_j^\dag a_j+a_j^\dag a_j)+{\rm irrelevant~terms}~~\label{4aa0}\\
A(j)^2&=&{1\over \omega_j}(2a_j^\dag a_j+1)+{\rm irrelevant~terms}
\ee
which lead to two relations that will be extensively used in later calculations\footnote{The reason of using $6\,A(i)^2A(j)^2$ instead of $A(i)^2A(j)^2$ is because that in the series expansion of the potential (\ref{potential}) it always appears  the combination factor $6\,A(i)^2A(j)^2$, as can be seen in several examples in next subsection.} 
\be
A(j)^4&=&6\times \left[{N_j+1\over \omega_j^2}a_j^\dag a_j\right]~~~\label{4aa}\\
6\,A(i)^2A(j)^2&=&6\times{1\over \omega_i\omega_j}\(4a_i^\dag a_ia_j^\dag a_j+2a_i^\dag a_i+2a_j^\dag a_j\)\nn
\\&=&6\times\left[{2N_i+2\over \omega_i\omega_j}a_j^\dag a_j+{2N_j+2\over \omega_i\omega_j}a_i^\dag a_i\right],~~~i\ne j~~~\label{2aa}
\ee
The term  $a_j^\dag a_ja_j^\dag a_j$  in (\ref{4aa0})  is written as $N_ja_j^\dag a_j$ in (\ref{4aa}),  as we did in sec. 3.2.   Sec.3.2 also tells us that we will let $N_j\rightarrow R_j^{(n-1)}N_j$ in calculating the complexity at the n'th  order of $\lambda$,

(III) Adopting  the series expansion (\ref{potential}), we can develop diagrammatic rules based on two basic elements, {\it ``circle''} and {\it ``pair''}, which appear in  (\ref{4aa}) and (\ref{2aa}).   We plot them in figure 1 and 2: 
\\
\\
\scalebox{0.15}{\hspace{36cm}\includegraphics{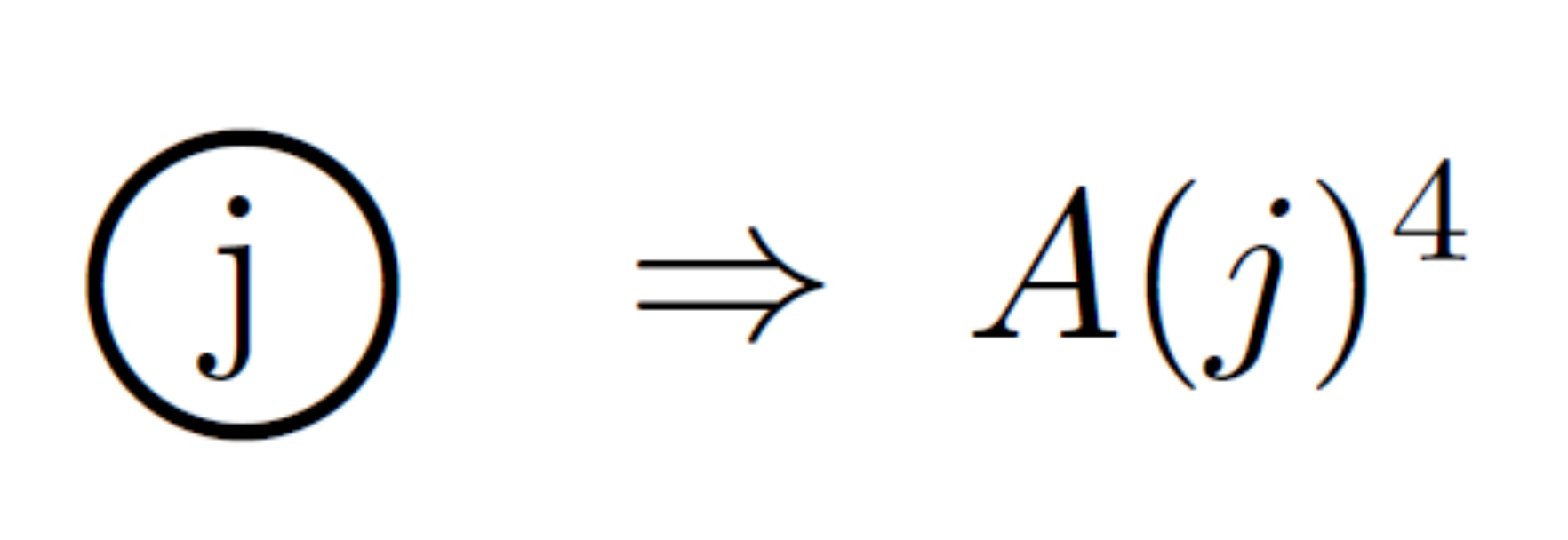}}
\\
{\color{white}0}\hspace{5cm}{Figure 1:  The {\it ``circle"} element}
\\
\\
\scalebox{0.18}{\hspace{28cm}\includegraphics{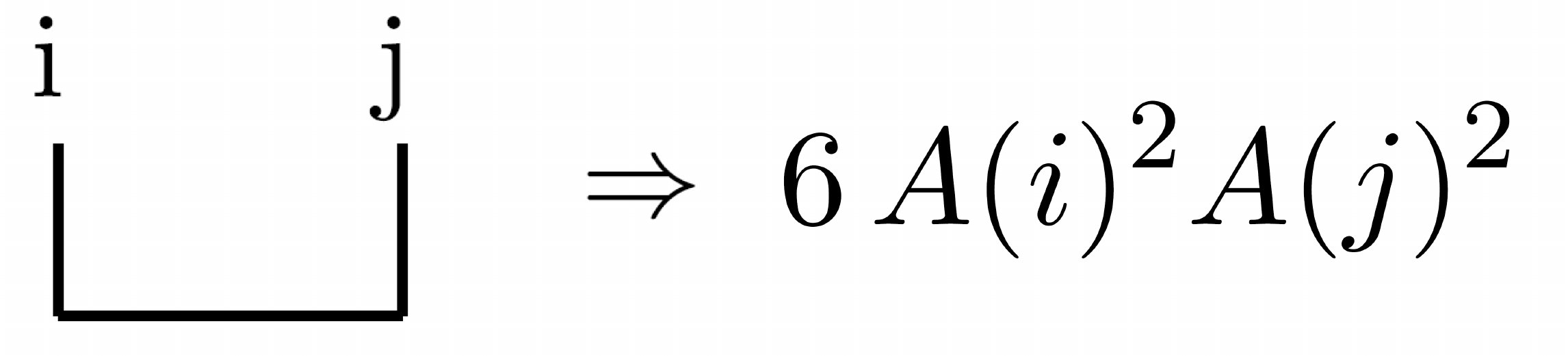}}
\\
{\color{white}0}\hspace{5cm}{ Figure 2:  The {\it ``pair"} element}

(IV) The diagrams for the potential $V$ then can be  classified into three classes: odd $N$, odd ${N\over 2}$, and even ${N\over 2}$.  We discuss corresponding rules in the following. 
\\

$\bullet$ Odd $N$  :  
We write numbers $1,2,....., N$ on a horizon  and  assign {\it ``circle"} on $N$. Then we assign {\it ``pair"} on $(1,N-1),...~(i,N-i),...,({N-1\over 2},{N+1\over 2})$.  

An N=9  example is plotted in figure 3.
\\
\scalebox{0.4}{\hspace{8cm}\includegraphics{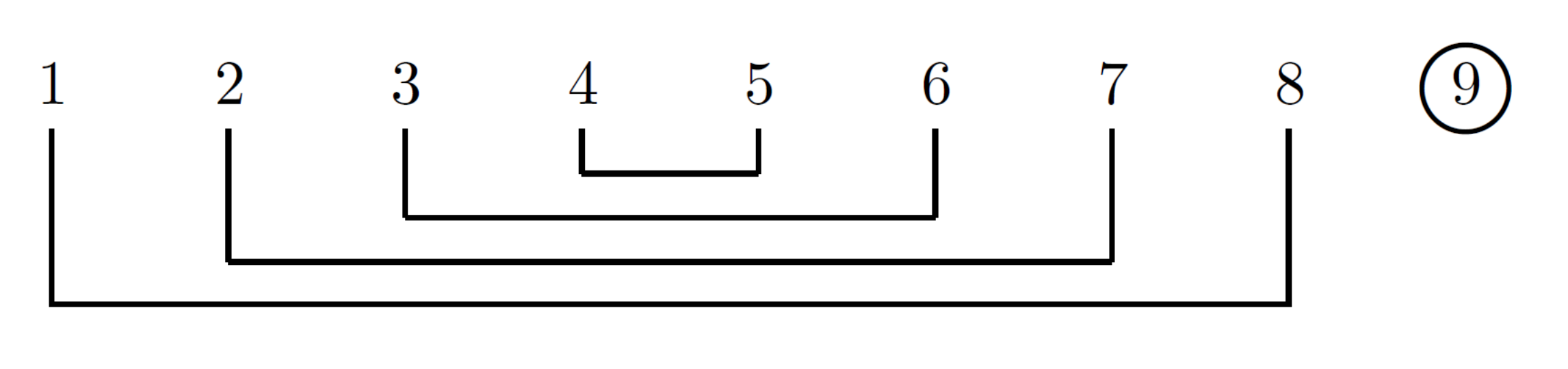}}
\\
{\color{white}0}\hspace{5cm}{Figure 3:  N=9 diagram}
\\

 $\bullet$ Odd ${N\over 2}$  :  
We write numbers $1,2,....., N$ on a horizon and   assign  {\it ``circle"} on ${N\over 2}$ and on ${N}$. 
We also assign a {\it ``pair"} on (${N\over 2}, N)$,  {\it ``pair"} on $(1,N-1),...~(i,N-i),...({N\over 2}-1,{N\over 2}+1)$, and assign {\it ``pair"} on $(1,{N\over 2}-1),...~(i,{N\over 2}-i),...,({N\over 4}-{1\over 2},{N\over 4}+{1\over 2})$.  Finally, we assign {\it ``pair"} on $({N\over 2}+1,N-1),.....~(i,{3N\over 2}-i),...,({3N\over 4}-{1\over 2},{3N\over 4}+{1\over 2})$. 

An N=10  example is plotted in figure 4.
\\
\scalebox{0.4}{\hspace{8cm}\includegraphics{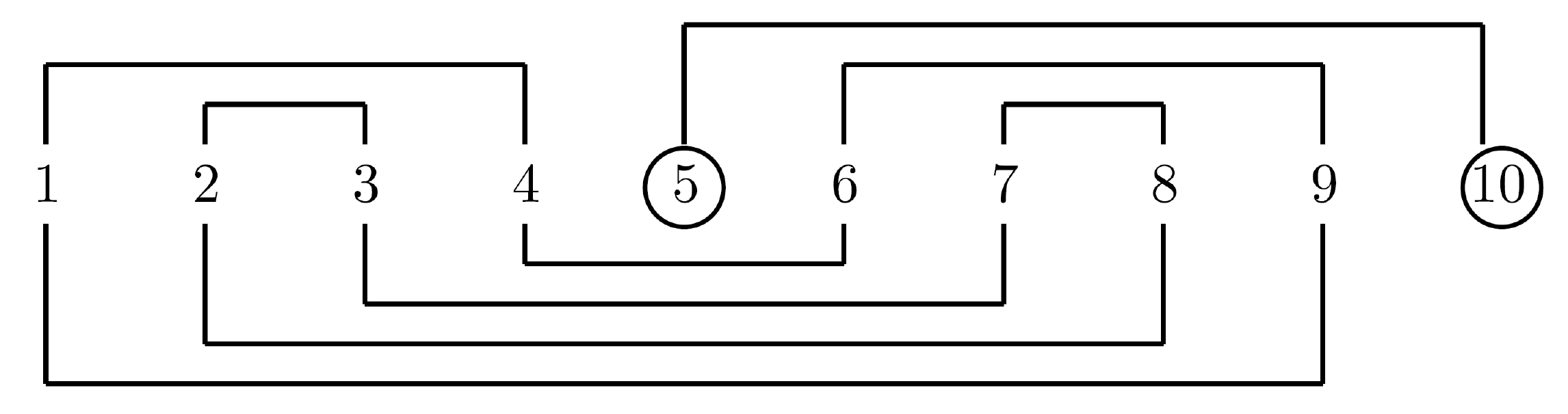}}
\\
{\color{white}0}\hspace{3cm}{ Figure 4:  N=10 diagram}
\\

 $\bullet$ Even ${N\over 2}$   :  
We write numbers $1,2,....., N$ on a horizon and  assign {\it ``circle"} on $N$, ${N\over 4}$, ${N\over 2}$, and ${3N\over 4}$.  
Also assign  {\it ``pair"} on (${N\over 2}, N)$ and (${N\over 4}, {3N\over 4})$. 
Assign {\it ``pair"} on $(1,N-1),...~(i,N-i),...({N\over 2}-1,{N\over 2}+1)$ and assign {\it ``pair"} on $(1,{N\over 2}-1),...~(i,{N\over 2}-i),...,({N\over 4}-1,{N\over 4}+1)$. 
Finally, we assign {\it ``pair"} on $({N\over 2}+1,N-1),.....~(i,{3N\over 2}-i),...,({3N\over 4}-1,{3N\over 4}+1)$

An N=12  example is plotted in figure 5.
\\
\scalebox{0.5}{\hspace{4cm}\includegraphics{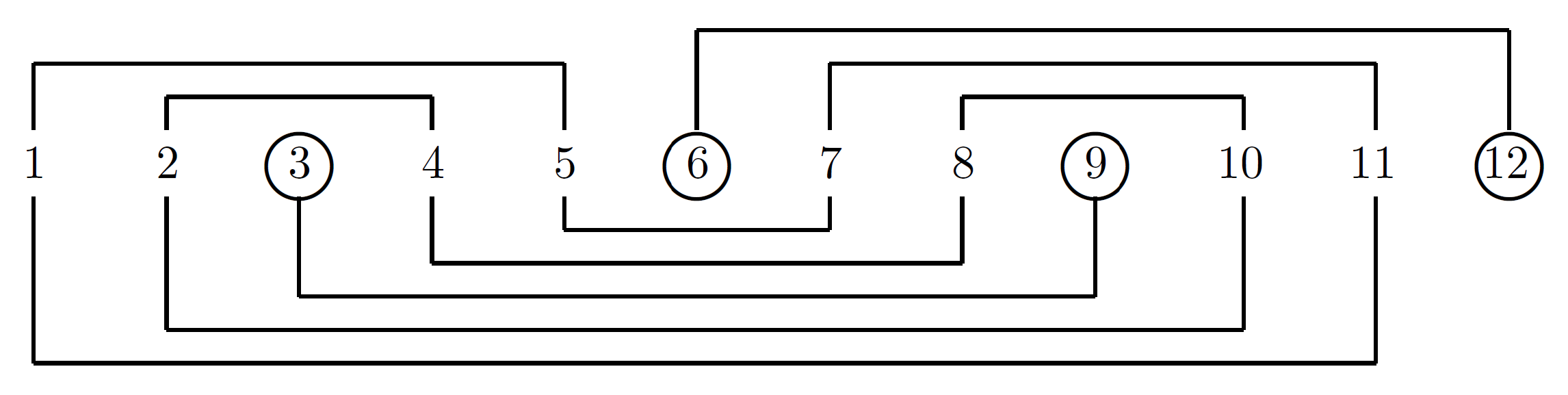}}
\\
{\color{white}0}\hspace{3cm}{Figure 5:  N=12 diagram}
\\

From figures 3, 4, and 5,  we can see a pairing property:  Assign the {\it ``circle"} element $A(j)^4$ pairing with ``j"  once and assign the {\it ``pair''} element $A(i)^2A(j)^2$ pairing with each ``i"  and ``j"  once, then  the odd $N$ diagrams have pairings in each ``j"  once while the even $N$ diagrams have pairings for each ``j" twice. 
\subsection{Interacting  Term of N Coupled Oscillators : Some Calculations }

We now take several values of N as examples to plot  the diagrams and use (\ref{4aa}) and (\ref{2aa}) to calculate  the associated complexity. General formulae will be presented in the next section.
\\

$\bullet$ N=2:  
\\
\scalebox{0.15}{\hspace{30cm}\includegraphics{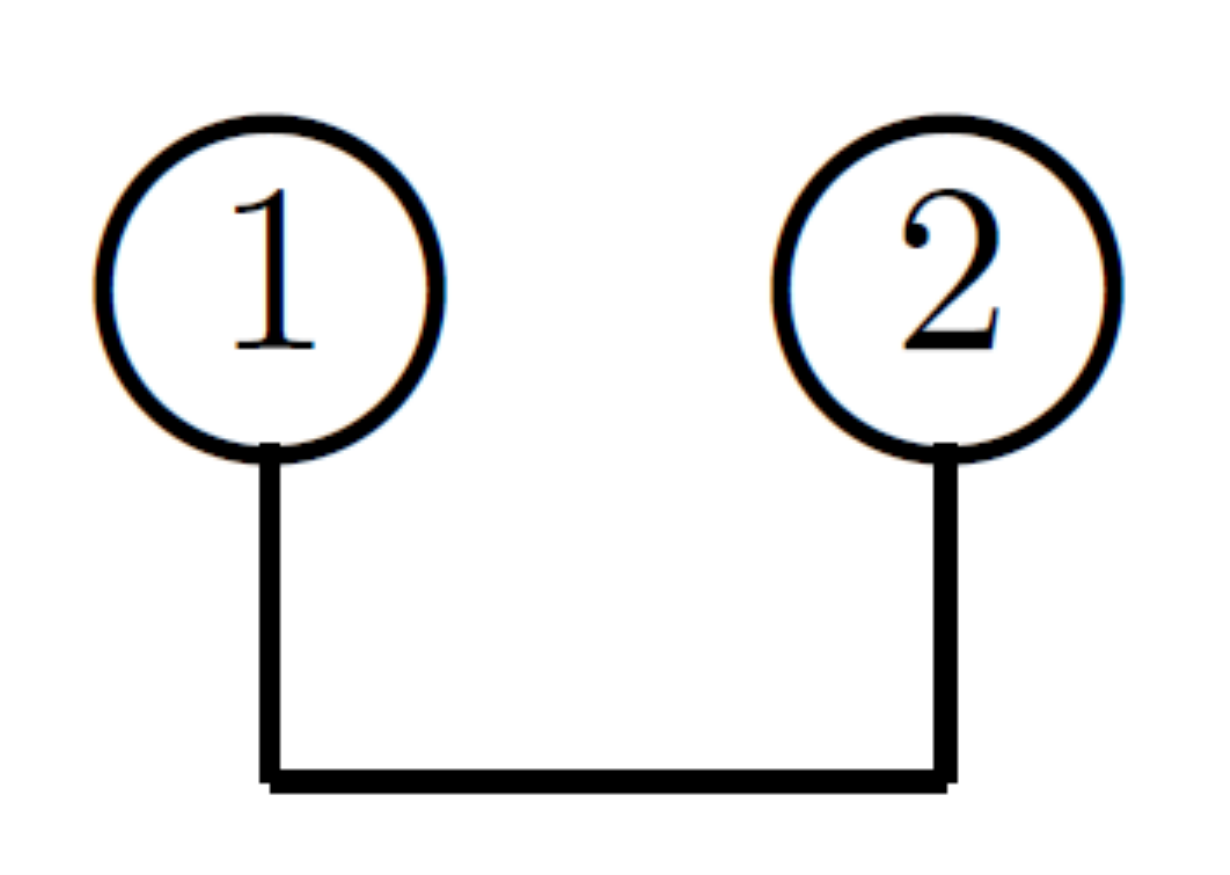}}
\\
{\color{white}0}\hspace{3cm}{Figure 6:  N=2 diagram}
\\
\\
As shown in figure 6 the series expansion (\ref{potential}) is 
\be
V_{N=2}&=& \frac{ \lambda}{4\cdot 2} \left[A(1)^4+A(2)^4+6 A(1)^2 A(2)^2\right]\nn\\
&=&\frac{6 \lambda}{4\cdot 2} \left[{1\over \omega_1^2}(N_1+1)a_1^\dag a_1+{1\over \omega_2^2}(N_2+1)a_2^\dag a_2+{2\over \omega_1\omega_2}\((N_2+1)a_1^\dag a_1+(N_1+1)a_2^\dag a_2\)
\right]\nn\\
&=&\frac{6 \lambda}{4\cdot 2} \left[\({1+N_1\over \omega_1^2}+{2+2N_2\over \omega_1\omega_2}\)a_1^\dag a_1+\({(1+N_2\over \omega_2^2}+{2+2N_1\over \omega_1\omega_2}\)a_2^\dag a_2 \ .
\right]
\ee
We have used (\ref{4aa}) and (\ref{2aa}).  The above result  matches with  (\ref{VN=2}).  

The associated complexity can be evaluated to any order in $\lambda$:
\be
R^{(n)}_1&=&{\omega_1+\frac{6 \lambda}{4\cdot 2}\,\({1+N_1R^{(n-1)}_1\over \omega_1^2}+ {2+2N_2R^{(n-1)}_2\over \omega_1\omega_2}\)\over  \omega_f+\frac{6 \lambda}{4\cdot 2}\,{1+N_1\over \omega_f^2}},~~
R^{(n)}_2={\omega_2+\frac{6 \lambda}{4\cdot 2}\,\({1+N_2R^{(n-1)}_2\over \omega_2^2}+ {2+2N_1R^{(n-1)}_1\over \omega_1\omega_2}\)\over  \omega_f+\frac{6 \lambda}{4\cdot 2}\,{2+2N_1\over \omega_f^2}}\nn\\
\ee
with initial values  $R^{(0)}_{(1,2)}$ defined in (\ref{R0}). For excited states, the $n$-order squared distance is $D^{(n)2}_{(N_1,N_2)}=\sum_{i=1}^2(N_i+1)\left(\ln \(\sqrt {R^{(n)}_i}\)\right)^2$, which is the $n$-order complexity of 2 coupled oscillators.  While above results exactly match (\ref{R2}) we have expressed them in the new form that helps us to identify rules for computing a general $N$ result.
\\

$\bullet$ N=3: 
\\
\scalebox{0.15}{\hspace{30cm}\includegraphics{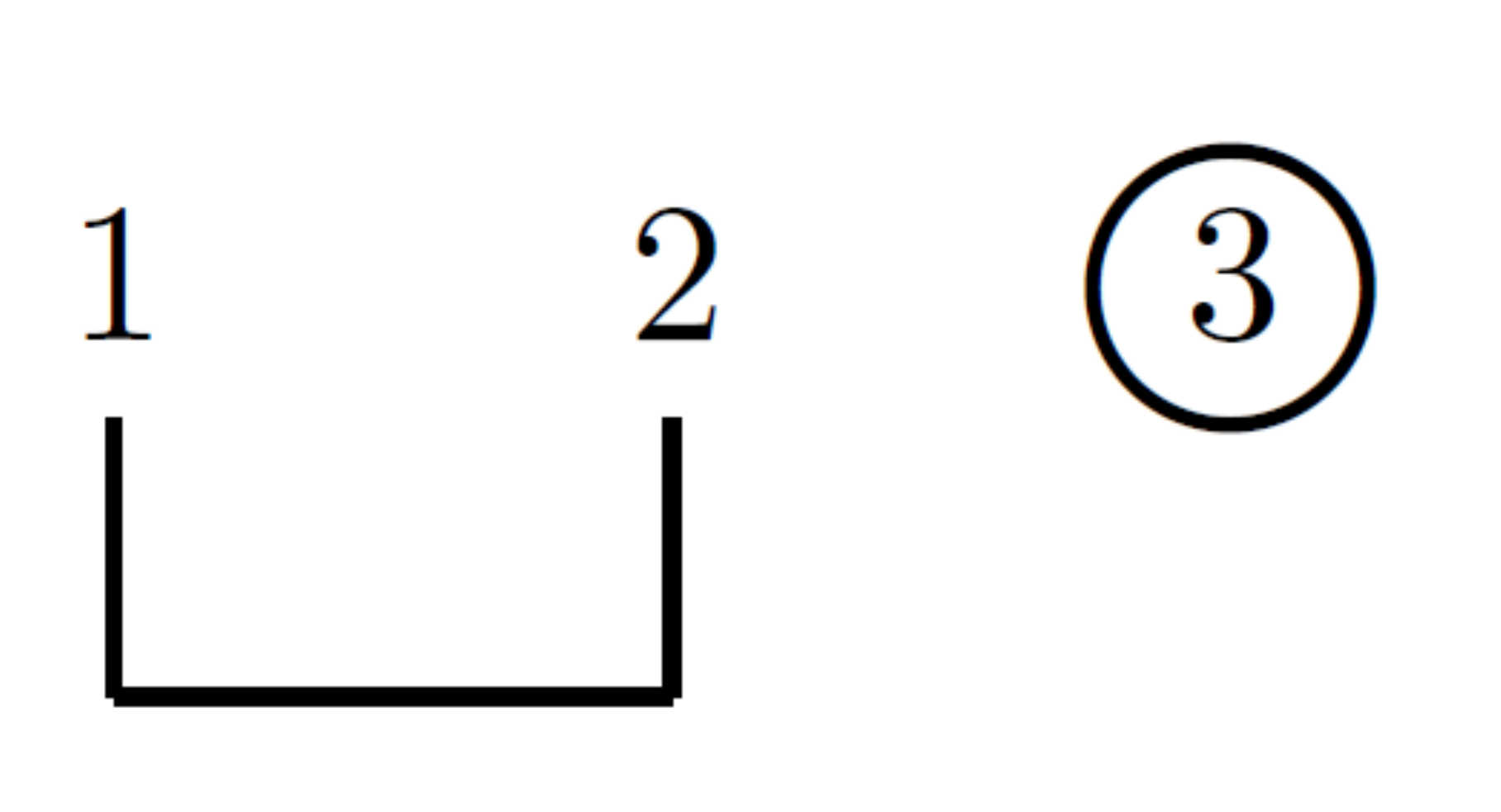}}
\\
{\color{white}0}\hspace{3.5cm}{Figure 7:  N=3 diagram}
\\
\\
As shown in figure 7 the series expansion (\ref{potential}) is
\be
 V_{N=3}&=&\frac{ \lambda}{4\cdot 3} \left[A(3)^4+6 A(1)^2 A(2)^2\right]\nn\\
&=&\frac{6 \lambda}{4\cdot 3} \left[{1+N_3\over \omega_3^2}a_3^\dag a_3+{2+2N_2\over \omega_1\omega_2}a_1^\dag a_1+{2+2N_1\over \omega_1\omega_2}a_2^\dag a_2\right]
\ee
We have recurrent relations
\be
R^{(n)}_1&=&{\omega_1+\frac{6 \lambda}{4\cdot 3}\,\({2+2N_2R^{(n-1)}_2\over \omega_1\omega_2}\)\over  \omega_f+\frac{6 \lambda}{4\cdot 3}\,\({2+2N_2\over \omega_f^2}\)},~~
R^{(n)}_2={\omega_2+\frac{6 \lambda}{4\cdot 3}\,\({2+2N_1R^{(n-1)}_1\over \omega_1\omega_2}\)\over  \omega_f+\frac{6 \lambda}{4\cdot 3}\,\({2+2N_1\over \omega_f^2}\)}\nn\\
R^{(n)}_3&=&{\omega_3+\frac{6 \lambda}{4\cdot 3}\,\({1+N_3R^{(n-1)}_3\over \omega_3^2}\)\over  \omega_f+\frac{6 \lambda}{4\cdot 3}\,\({1+N_3\over \omega_f^2}\)} \ .
\ee
For excited states, $ D^{(n)2}_{(N_1,N_2,N_3)}=\sum_{i=1}^3(N_i+1)\left(\ln \(\sqrt {R^{(n)}_i}\)\right)^2$, which is the $n$-order complexity of 3 coupled oscillators.
\\

$\bullet$ N=4: 
\\
\scalebox{0.15}{\hspace{46cm}\includegraphics{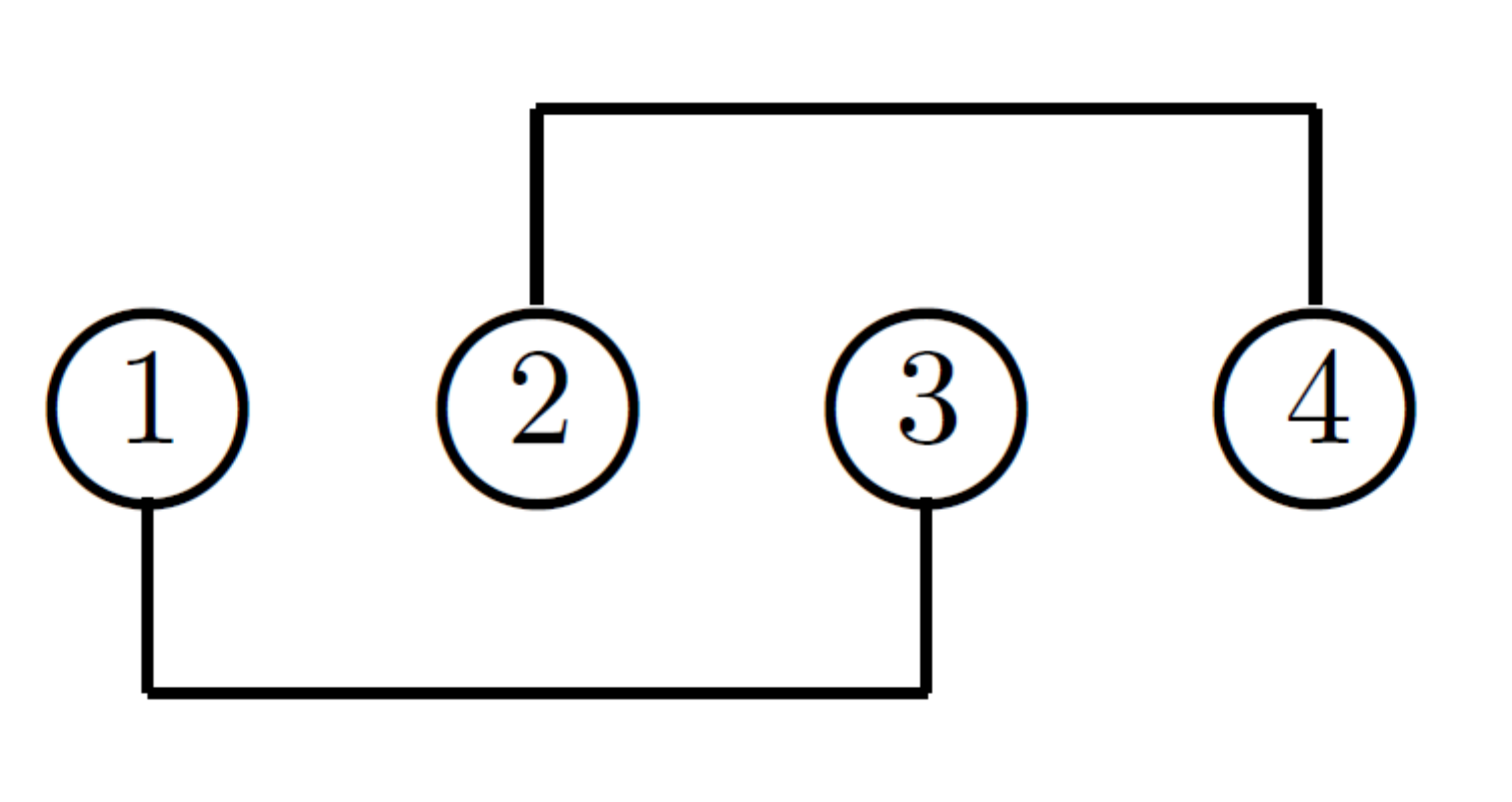}}
\\
{\color{white}0}\hspace{6cm}{Figure 8:  N=4 diagram}
\\
\\
As shown in figure 8 the series expansion (\ref{potential}) is 
\be
 V_{N=4}&=&\frac{\lambda}{4\cdot 4} \left[A(1)^4+A(2)^4+A(3)^4+A(4)^4+6  A(1)^2A(3)^2+6 A(2)^2 A(4)^2\right]\nn\\
&=&\frac{6 \lambda}{4\cdot4} \left[\({1+N_1\over \omega_1^2}+{2+2N_3\over \omega_1\omega_3}\)a_1^\dag a_1+\({1+N_2\over \omega_2^2}+{2+2N_4\over \omega_2\omega_4}\)a_2^\dag a_2 \right.\nn
\\
&&\left.~~+\({1+N_3\over \omega_3^2}+{2+2N_1\over \omega_1\omega_3}\)a_3^\dag a_3+\({1+N_4\over \omega_4^2}+{2+2N_2\over \omega_2\omega_4}\)a_4^\dag a_4 \right] \ ,
\ee
We have recurrent relations
\be
R^{(n)}_1&=&{\omega_1+\frac{6 \lambda}{4\cdot 4}\,\({1+N_1R^{(n-1)}_1\over \omega_1^2}+ {2+2N_3R^{(n-1)}_3\over \omega_1\omega_3}\)\over  \omega_f+\frac{6 \lambda}{4\cdot 4}\,{1+N_1\over \omega_f^2}},~~
R^{(n)}_2={\omega_2+\frac{6 \lambda}{4\cdot 4}\,\({1+N_2R^{(n-1)}_2\over \omega_2^2}+ {2+2N_4R^{(n-1)}_4\over \omega_2\omega_4}\)\over  \omega_f+\frac{6 \lambda}{4\cdot 4}\,{2+2N_4\over \omega_f^2}}\nn\\
\nn\\
R^{(n)}_3&=&{\omega_3+\frac{6 \lambda}{4\cdot 4}\,\({1+N_3R^{(n-1)}_3\over \omega_3^2}+ {2+2N_1R^{(n-1)}_1\over \omega_1\omega_3}\)\over  \omega_f+\frac{6 \lambda}{4\cdot 4}\,{1+N_3\over \omega_f^2}},~~
R^{(n)}_4={\omega_4+\frac{6 \lambda}{4\cdot 4}\,\({1+N_4R^{(n-1)}_4\over \omega_4^2}+ {2+2N_2R^{(n-1)}_2\over \omega_2\omega_4}\)\over  \omega_f+\frac{6 \lambda}{4\cdot 4}\,{1+N_4\over \omega_f^2}} \  . \nn\\
\ee
 For excited state, $
D^{(n)2}_{(N_1,N_2,N_3,N_4)}=\sum_{i=1}^4(N_i+1)\left(\ln \(\sqrt {R^{(n)}_i}\)\right)^2$ , which is the $n$-order complexity of 4 coupled oscillators.
\\

$\bullet$ N=5: 
\\
\scalebox{0.2}{\hspace{20cm}\includegraphics{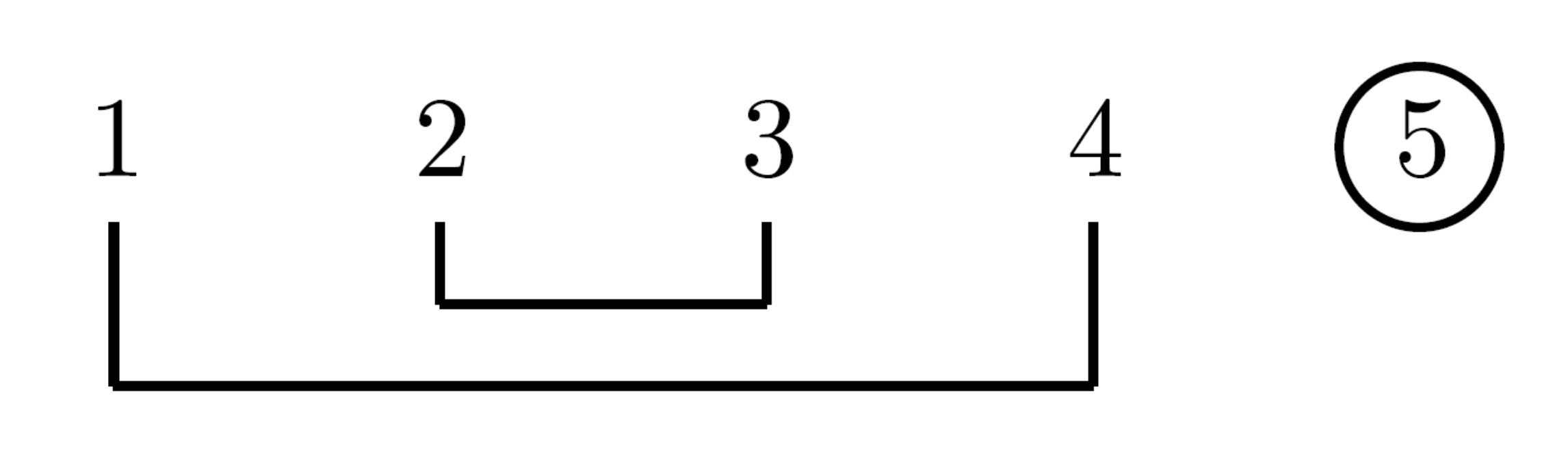}}
\\
{\color{white}0}\hspace{4cm}{Figure 9:  N=5 diagram}
\\
As shown in figure 9 the series expansion (\ref{potential}) is 
\be
V_{N=5}&=&\frac{ \lambda}{4\cdot 5} \left[6 A(1)^2 A(4)^2+6 A(2)^2 A(3)^2+A(5)^4\right]\\
&=&\frac{6 \lambda}{4\cdot 3} \left[{1+N_5\over \omega_5^2}a_5^\dag a_5+{2+2N_4\over \omega_1\omega_4}a_1^\dag a_1+{2+2N_3\over \omega_3\omega_2}a_2^\dag a_2+{2+2N_2\over \omega_3\omega_2}a_3^\dag a_3+{2+2N_1\over \omega_1\omega_4}a_4^\dag a_4\right]\nn\\
\ee
We have recurrent relations
\be
R^{(n)}_1&=&{\omega_1+\frac{6 \lambda}{4\cdot 3}\,\({2+2N_4R^{(n-1)}_4\over \omega_1\omega_4}\)\over  \omega_f+\frac{6 \lambda}{4\cdot 3}\,\({2+2N_4\over \omega_f^2}\)},~~
R^{(n)}_2={\omega_2+\frac{6 \lambda}{4\cdot 3}\,\({2+2N_3R^{(n-1)}_3\over \omega_3\omega_2}\)\over  \omega_f+\frac{6 \lambda}{4\cdot 3}\,\({2+2N_3\over \omega_f^2}\)}\nn\\
R^{(n)}_3&=&{\omega_3+\frac{6 \lambda}{4\cdot 3}\,\({2+2N_2R^{(n-1)}_2\over \omega_3\omega_2}\)\over  \omega_f+\frac{6 \lambda}{4\cdot 3}\,\({2+2N_2\over \omega_f^2}\)},~~
R^{(n)}_4={\omega_4+\frac{6 \lambda}{4\cdot 3}\,\({2+2N_1R^{(n-1)}_1\over \omega_1\omega_4}\)\over  \omega_f+\frac{6 \lambda}{4\cdot 3}\,\({2+2N_1\over \omega_f^2}\)}\nn\\
R^{(n)}_5&=&{\omega_5+\frac{6 \lambda}{4\cdot 3}\,\({1+N_5R^{(n-1)}_5\over \omega_5^2}\)\over  \omega_f+\frac{6 \lambda}{4\cdot 3}\,\({1+N_5\over \omega_f^2}\)} \ .
\ee
For excited states, $ D^{(n)2}_{(N_1,N_2,N_3)}=\sum_{i=1}^5(N_i+1)\left(\ln \(\sqrt {R^{(n)}_i}\)\right)^2$, which is the $n$-order complexity of 5 coupled oscillators.
\\

$\bullet$ N=6: 
\\
\scalebox{0.25}{\hspace{16cm}\includegraphics{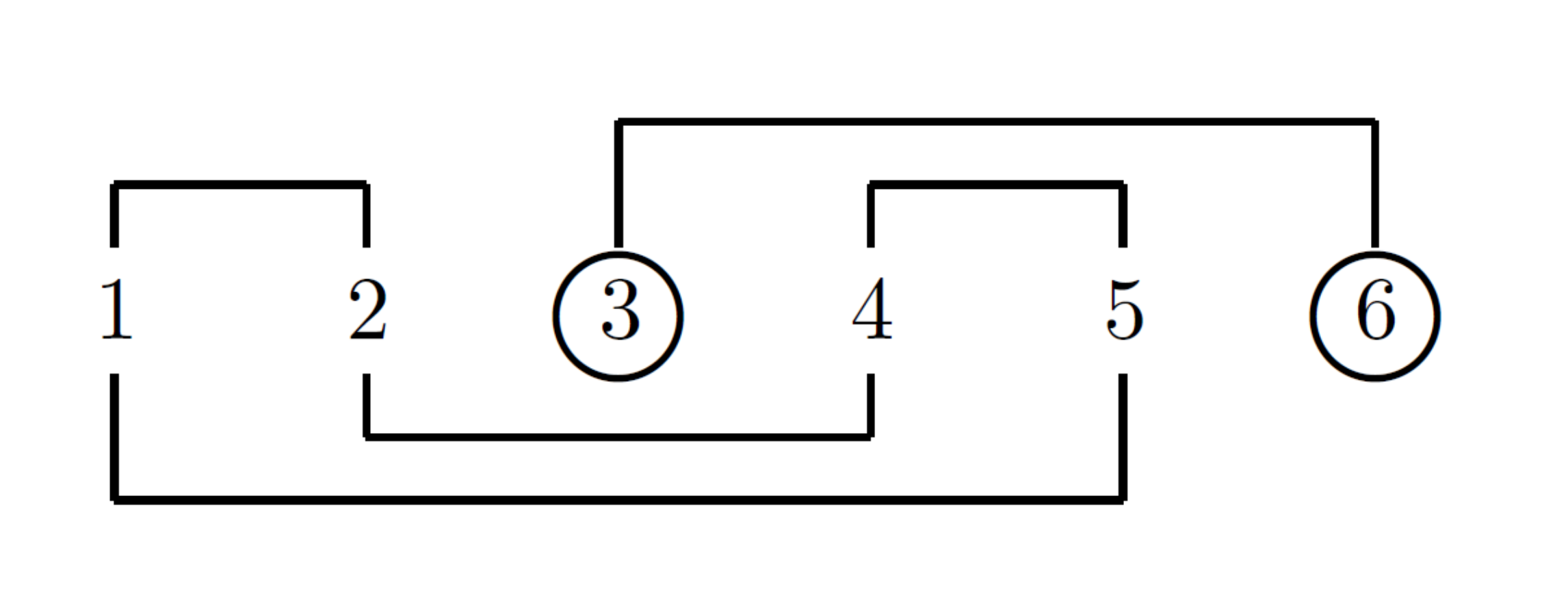}}
\\
{\color{white}0}\hspace{4cm}{Figure 10:  N=6 diagram}
\\
As shown in figure 10 the series expansion (\ref{potential}) is 
\be
V_{N=6}&=&\frac{ \lambda}{4\cdot 6} \left[A(3)^4+A(6)^4+6A(1)^2A(5)^2 +6A(2)^2A(4)^2+6A(1)^2A(2)^2\right.\nn\\
&&~~~~~\left.+6A(3)^2A(6)^2+6A(4)^2A(5)^2\right]
\ee

$\bullet$ N=7: 
\\
\scalebox{0.25}{\hspace{20cm}\includegraphics{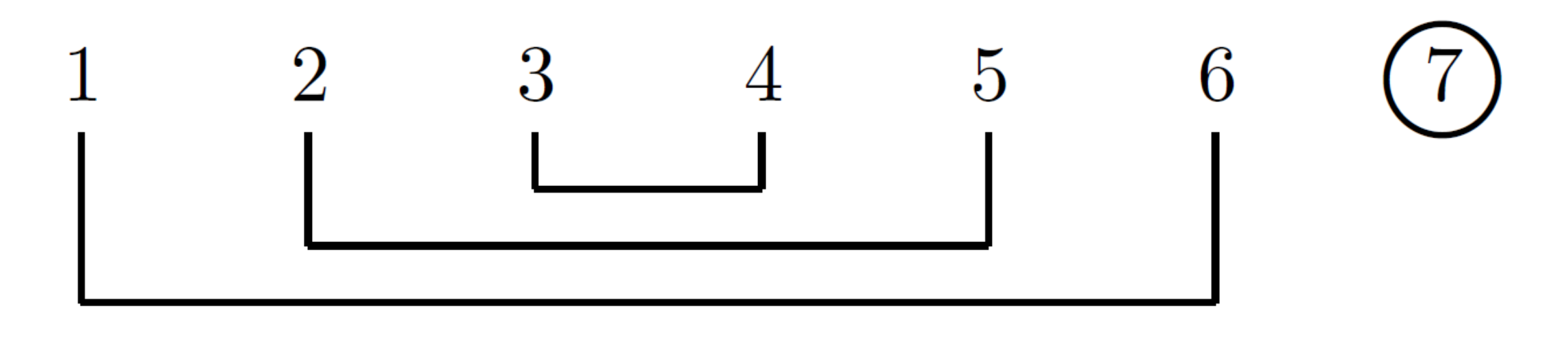}}
\\
{\color{white}0}\hspace{5cm}{Figure 11:  N=7 diagram}
\\
As shown in figure 11 the series expansion (\ref{potential}) is 
\be
V_{N=7}=\frac{\lambda}{4\cdot 7} \left[6 A(1)^2 A(6)^2+6 A(2)^2 A(5)^2+6 A(3)^2 A(4)^2+A(7)^4\right]
\ee

$\bullet$ N=8: 
\\
\scalebox{0.25}{\hspace{20cm}\includegraphics{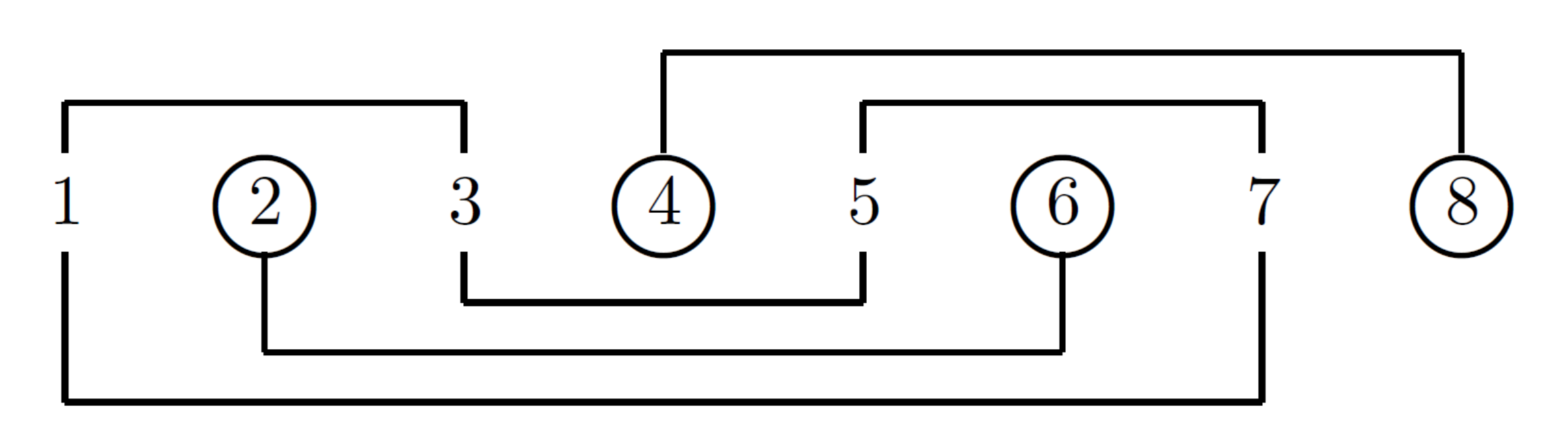}}
\\
{\color{white}0}\hspace{5cm}{Figure 12:  N=8 diagram}
\\
As shown in figure 12 the series expansion (\ref{potential}) is 
\be
V_{N=8}&=&\frac{\lambda}{4\cdot 8} \left[A(2)^4+A(4)^4+A(6)^4+A(8)^4+6 A(1)^2 A(7)^2+6 A(2)^2 A(6)^2+6 A(3)^2 A(5)^2\right.\nn\\
&&\left.+6A(1)^2A(3)^2+6A(4)^2A(8)^2+6A(5)^2A(7)^2\right] 
\ee
With  these experiences we will in the next section derive general formulae of the complexity for any $N$ to any order in $\lambda$.

\section{Complexity of N Coupled Oscillators}
\subsection{Complexity of N Coupled Oscillators : General Formulae}
From the above analysis and  relations (\ref{4aa}) and (\ref{2aa}), we find 
\be
V_N&=&\frac{\lambda}{4\cdot N} \left[\sum_{\rm ``cirle''\,j}A(j)^4+6\sum_{\rm ``pair''\,(i,j)}A(i)^2A(j)^2\right]\nn\\
&=&\frac{6\lambda}{4\cdot N} \left[\sum_{\rm ``cirle''\,i}{(1+N_j)\over \omega_j^2}a_j^\dag a_j+\sum_{\rm ``pair''\,(i,j)}{2+2N_i\over \omega_i\omega_j}a_j^\dag a_j+{2+2N_j\over \omega_i\omega_j}a_i^\dag a_i\right]~~\label{VN}
\ee
where {\it ``cirle''} and {\it ``pair''} can be read from diagrams; see figures 3, 4, and 5. 
\\

$\bullet$ Odd $N$ : We recall, from Sec 4.2, the odd N case is simplest as it only has one {\it ``circle"} located at $N$, and each {\it ``pair"} is independent to each other (figure 3).  
Eq(\ref{VN}) becomes
\be
V_{\rm odd\,N}&=&\frac{6\lambda}{4\cdot N} \left[{1+N_N\over \omega_N^2}a_N^\dag a_N+\sum_{i=1}^{N-1}{2+2N_{N-i}\over \omega_i\omega_{N-i}}a_{i}^\dag a_{i}\right]
\ee
By adding  the kinematic term (\ref{K}) and defining the recursion relations
\be
R_{N,odd}^{(n)}&=&{{\omega_N}+\frac{6\lambda}{4\cdot N} {1+N_NR_N^{(n-1)}\over \omega_N^2}\over 
{\omega_f}+\frac{6\lambda}{4\cdot N} {1+N_N\over \omega_f^2}}\\
R_{i,odd}^{(n)}&=&{{\omega_i}+\frac{6\lambda}{4\cdot N} {2+2N_{N-i}R_{N-i}^{(n-1)}\over \omega_i\omega_{N-i}}\over {\omega_f}+\frac{6\lambda}{4\cdot N} {2+2N_{N-i}\over \omega_f^2}},~~~1\le i\le N-1  \ , 
\ee
 the n-order complexity is
\be
 D^{(n)2}_{(N_1,...,N_N)}=(N_N+1)\left(\ln \(\sqrt {R^{(n)}_{N,odd}}\)\right)^2+\sum_{i=1}^{N-1}(N_i+1)\left(\ln \(\sqrt {R^{(n)}_{i,odd}}\)\right)^2 \ ,~~\label{F1}
\ee
where $R^{(0)}_i$ is defined in  (\ref{R0}).
\\

$\bullet$ Odd ${N\over 2}$ :  These cases have two {\it ``circle"} located at ${N\over 2}$ and $N$, pairing with each other (figure 4).  The potential is
\be
V_{\rm odd\,{N\over 2}}^{\it ``circle"}&=&\frac{6\lambda}{4\cdot N} \left[\left({1+N_{N\over 2}\over \omega_{N\over 2}^2}+{2+2N_{N}\over \omega_{N\over 2}\omega_{N}}\right)a_{N\over 2}^\dag a_{N\over 2}+\left({1+N_{N}\over \omega_{N}^2}+{2+2N_{N\over 2}\over \omega_{N\over 2}\omega_{N}}\right)a_{N}^\dag a_{N}\right] 
\ee
The remaining contributions are those from pure {\it ``pairing''} sites.   Recalling the figure 2 and the relation (\ref{2aa}) we can evaluate the corresponding  potential. 
The result is
\be
V_{\rm odd\,{N\over 2}}^{\it ``pair''}&=&\frac{6\lambda}{4\cdot N} \left[\sum_{i=1}^{{N\over 2}-1}\({2+2N_{N-i}\over \omega_i\omega_{N-i}}+{2+2N_{{N\over 2}-i}\over \omega_i\omega_{{N\over 2}-i}}\)a_{i}^\dag a_{i}    \right. \nn\\
&&~~~~~~~~~~\left.\sum_{i={N\over 2}+1}^{N-1}\({2+2N_{N-i}\over \omega_i\omega_{N-i}}+{2+2N_{{3N\over 2}-i}\over \omega_i\omega_{{3N\over 2}-i}}\)a_{i}^\dag a_{i} \right]~~~\label{VPO}
\ee
By adding  the kinematic term (\ref{K}) and defining the recursion relations
\be
R_{{N\over 2},even}^{(n)}&=&{{\omega_{N\over 2}}+\frac{6\lambda}{4\cdot N}\({1+N_{N\over 2}R_{N\over 2}^{(n-1)}\over \omega_{N\over 2}^2}+{2+2N_{N}R_{N}^{(n-1)}\over \omega_{N\over 2}\omega_{N}}\)
\over {\omega_f}+\frac{6\lambda}{4\cdot N}\,{1+N_{N\over 2}\over \omega_{0}^2}}\ ,
\ee
\be
R_{N,even}^{(n)}&=&{{\omega_N}+\frac{6\lambda}{4\cdot N}\({1+N_{N}R_{N}^{(n-1)}\over \omega_{N}^2}+{2+2N_{N\over 2}R_{N\over 2}^{(n-1)}\over \omega_{N\over 2}\omega_{N}}\)
\over {\omega_f}+\frac{6\lambda}{4\cdot N}\,{1+N_{N}\over \omega_{0}^2}} \ , \\
\nn\\
R_{i,even}^{(n)}&=&{{\omega_i}+\frac{6\lambda}{4\cdot N} ({2+2N_{N-i}R_{N-i}^{(n-1)}\over \omega_i\omega_{N-i}}+{2+2N_{{N\over 2}-i}R_{{N\over 2}-i}^{(n-1)}\over \omega_i\omega_{{N\over 2}-i}}\)\over {\omega_f}+\frac{6\lambda}{4\cdot N} \,{2+2N_{N-i}\over\omega_{0}^2}},~~~1\le i\le {N\over 2}-1 \ , \\
\nn\\
\tilde R_{i,even}^{(n)}&=&{{\omega_i}+\frac{6\lambda}{4\cdot N} ({2+2N_{N-i}R_{N-i}^{(n-1)}\over \omega_i\omega_{N-i}}+{2+2N_{{3N\over 2}-i}R_{{3N\over 2}-i}^{(n-1)}\over \omega_i\omega_{{3N\over 2}-i}}\)\over {\omega_f}+\frac{6\lambda}{4\cdot N}\, {2+2N_{{3N\over 2}-i}\over \omega_{0}^2}},~~~{N\over 2}+1\le i\le N-1 \ ,
\ee
the $n$-order complexity is
\be
 D^{(n)2}_{(N_1,...,N_N)}&=&(N_{N\over2}+1)\left(\ln \(\sqrt {R^{(n)}_{{N\over2},even}}\)\right)^2+(N_N+1)\left(\ln \(\sqrt {R^{(n)}_{N,even}}\)\right)^2\nn\\
&&+\sum_{i=1}^{{N\over2}-1}(N_i+1)\left(\ln \(\sqrt {R^{(n)}_{i,even}}\)\right)^2
+\sum_{i={N\over2}+1}^{N-1}(N_i+1)\left(\ln \(\sqrt {\tilde R^{(n)}_{i,even}}\)\right)^2~~\label{F2}
\ee
where $R^{(0)}_i$ is defined in  (\ref{R0}).\\

$\bullet$ Even ${N\over 2}$:  These cases have two {\it ``circle"}  locate at ${N\over 2}$ and $N$, pairing with each other,  and two {\it ``circle"} locate at ${N\over 4}$ and ${3N\over 4}$, pairing with each other as well (figure 5).  
The potential is 
\be
V_{\rm even\,{N\over 2}}^{\it ``circle"}&=&\frac{6\lambda}{4\cdot N} \left[\left({1+N_{N\over 2}\over \omega_{N\over 2}^2}+{2+2N_{N}\over \omega_{N\over 2}\omega_{N}}\right)a_{N\over 2}^\dag a_{N\over 2}+\left({1+N_{N}\over \omega_{N}^2}+{2+2N_{N\over 2}\over \omega_{N\over 2}\omega_{N}}\right)a_{N}^\dag a_{N}\right.\nn\\
&&~~\left.   + \left({1+N_{N\over 4}\over \omega_{N\over4}^2}+{2+2N_{3N\over 4}\over \omega_{N\over 4}\omega_{3N\over 4}}\right)a_{N\over 4}^\dag a_{N\over 4}+ \left({1+N_{3N\over 4}\over \omega_{3N\over4}^2}+{2+2N_{N\over 4}\over \omega_{3N\over 4}\omega_{N\over 4}}\right)a_{N\over 4}^\dag a_{3N\over 4}\right] ~~~
\ee
Again, the remaining contributions are those from pure {\it ``pairing''} sites.  We find 
\be
V_{\rm even\,{N\over 2}}^{\it ``pair''}&=&\frac{6\lambda}{4\cdot N} \left[\sum_{i=1,\ne{N\over 4}}^{{N\over 2}-1}\({2+2N_{N-i}\over \omega_i\omega_{N-i}}+{2+2N_{{N\over 2}-i}\over \omega_i\omega_{{N\over 2}-i}}\)a_{i}^\dag a_{i}    \right.\nn\\
\nn\\
&&~~~~~~~~~~\left.\sum_{i={N\over 2}+1,\ne{3N\over 4}}^{N-1}\({2+2N_{N-i}\over \omega_i\omega_{N-i}}+{2+2N_{{3N\over 2}-i}\over \omega_i\omega_{{3N\over 2}-i}}\)a_{i}^\dag a_{i} \right]
\ee
The above result  is the same as the odd ${N\over 2}$, i.e. (\ref{VPO}), but  drop the {\it ``circle"} at ${N\over 4}$ and  ${3N\over 4}$ since the potential of the two {\it ``circle"} has been considered in $V_{\rm even\,{N\over 2}}^{\it ``circle"}$.  

 By adding  the kinematic term (\ref{K}) and defining the recursion relations
\be
R_{{N\over 4},even}^{(n)}&=&{{\omega_{N\over 4}}+\frac{6\lambda}{4\cdot N}\({1+N_{N\over 4}R_{N\over 4}^{(n-1)}\over \omega_{N\over 4}^2}+{2+2N_{3N\over 4}R_{3N\over 4}^{(n-1)}\over \omega_{N\over 4}\omega_{3N\over 4}}\)
\over {\omega_f}+\frac{6\lambda}{4\cdot N}\,{1+N_{N\over 4}\over \omega_{0}^2}} \ , \\
\nn\\
R_{{3N\over 4},even}^{(n)}&=&{{\omega_{3N\over 4}}+\frac{6\lambda}{4\cdot N}\({1+N_{{3N\over 4}}R_{{3N\over 4}}^{(n-1)}\over \omega_{{3N\over 4}}^2}+{2+2N_{N\over 4}R_{N\over 4}^{(n-1)}\over \omega_{N\over 4}\omega_{{3N\over 4}}}\)
\over {\omega_f}+\frac{6\lambda}{4\cdot N}\,{2+2N_{N\over 4}\over \omega_{0}^2}} \ , 
\ee
the n-order complexity is
\be
 D^{(n)2}_{(N_1,...,N_N)}&=&(N_{N\over2}+1)\left(\ln \(\sqrt {R^{(n)}_{{N\over2},even}}\)\right)^2+(N_N+1)\left(\ln \(\sqrt {R^{(n)}_{N,even}}\)\right)^2\nn\\
&&+(N_{N\over4}+1)\left(\ln \(\sqrt {R^{(n)}_{{N\over4},even}}\)\right)^2+(N_{3N\over4}+1)\left(\ln \(\sqrt {R^{(n)}_{{3N\over4},even}}\)\right)^2\nn\\
&&+\sum_{i=1,\ne {N\over 4}}^{{N\over2}-1}(N_i+1)\left(\ln \(\sqrt {R^{(n)}_{i,even}}\)\right)^2
+\sum_{i={N\over2}+1,\ne {3N\over 4}}^{N-1}(N_i+1)\left(\ln \(\sqrt {\tilde R^{(n)}_{i,even}}\)\right)^2  \ . ~~\label{F3}~~~ 
\ee

These general formulae allow one  to  obtain higher-order complexity for excited states at any $N$ coupled oscillators, which is a lattice version of $\lambda \phi^4$ theory. 

While the  formulas we used involving the bare parameters to extract physics we have to rewrite the expressions above in terms of the renormalized quantities.  The issues had been studied in  \cite{Bhattacharyya1880} and we summarize it in below. First,  following\cite{smit} the  the mass have relation
 \be
 (m\,\delta)^2= (m_{R}\,\delta)^2-\frac{ \lambda_{R}\,\delta^{4-d}}{2}I (m_R\delta)+O(\lambda_R^2)\,.
 \ee
 where $m_R$ is the renormalized mass and $\lambda_R$ is the renormalized coupling defined at zero momentum. A running renormalized coupling can also be defined at finite momentum $\mu$ and for the leading order in the coupling one can replace $\lambda_R$ by $\lambda_R(\mu)$. Here,
 \be
 I(m_{R}\delta)=\prod_{i=1}^{d}\Big[\int_{-\pi}^{\pi}\frac{dl_i}{(2\pi)}\Big]\,\frac{1}{(m_R\,\delta)^2+4\,\sum_{i=1}^{d}\sin^2(\frac{l_i}{2})}
 \ee
 For $d=2$ 
 \be
 (m \,\delta)^2=(m_{R}\,\delta)^2-\frac{ \lambda_{R}\,\delta^2}{2}\left[C_0-2\,C_1 \log ( m_{R}\,\delta)- C_2 (m_{R}\,\delta)^2+\frac{1}{32\,\pi}(m_R\,\delta)^2\log ( (m_R\,\delta)^2)+\mathcal{O}((m_{R}\,\delta)^4)\right].\nn\\
 \ee
Here, $C_0= 0.28, C_1=0.08,C_2=0.02.$ For  $d\geq 3$ 
 \be
(m \,\delta)^2=(m_{R}\,\delta)^2-\frac{\lambda_R\,\delta^{4-d}}{2}\left[C_0- C_2 (m_{R}\,\delta)^2+\frac{1}{16 \pi^2} (m_{R}\,\delta)^2\log ((m_{R}\,\delta)^2)|_{d=4}+\mathcal{O}((m_{R}\,\delta)^4)\right].
\ee 
in which for $d=4$ there is an extra log term. The values of $C_0$ and $C_2$ for various dimensions can be found in \cite{Bhattacharyya1880}. Note that in leading order $\hat \lambda_0=\lambda_{R},$ where $\lambda_R$ is the renormalized coupling.

\subsection{Complexity of N Coupled Oscillators : Numerical Results}
We now use above formulas to perform numerical calculations and plot  several diagrams to  illuminate the properties of complexity. 
\\

(1) We plot in figure 13 the complexity for  various  lattice site number $N$. 
\\
\scalebox{0.5}{\hspace{6cm}\includegraphics{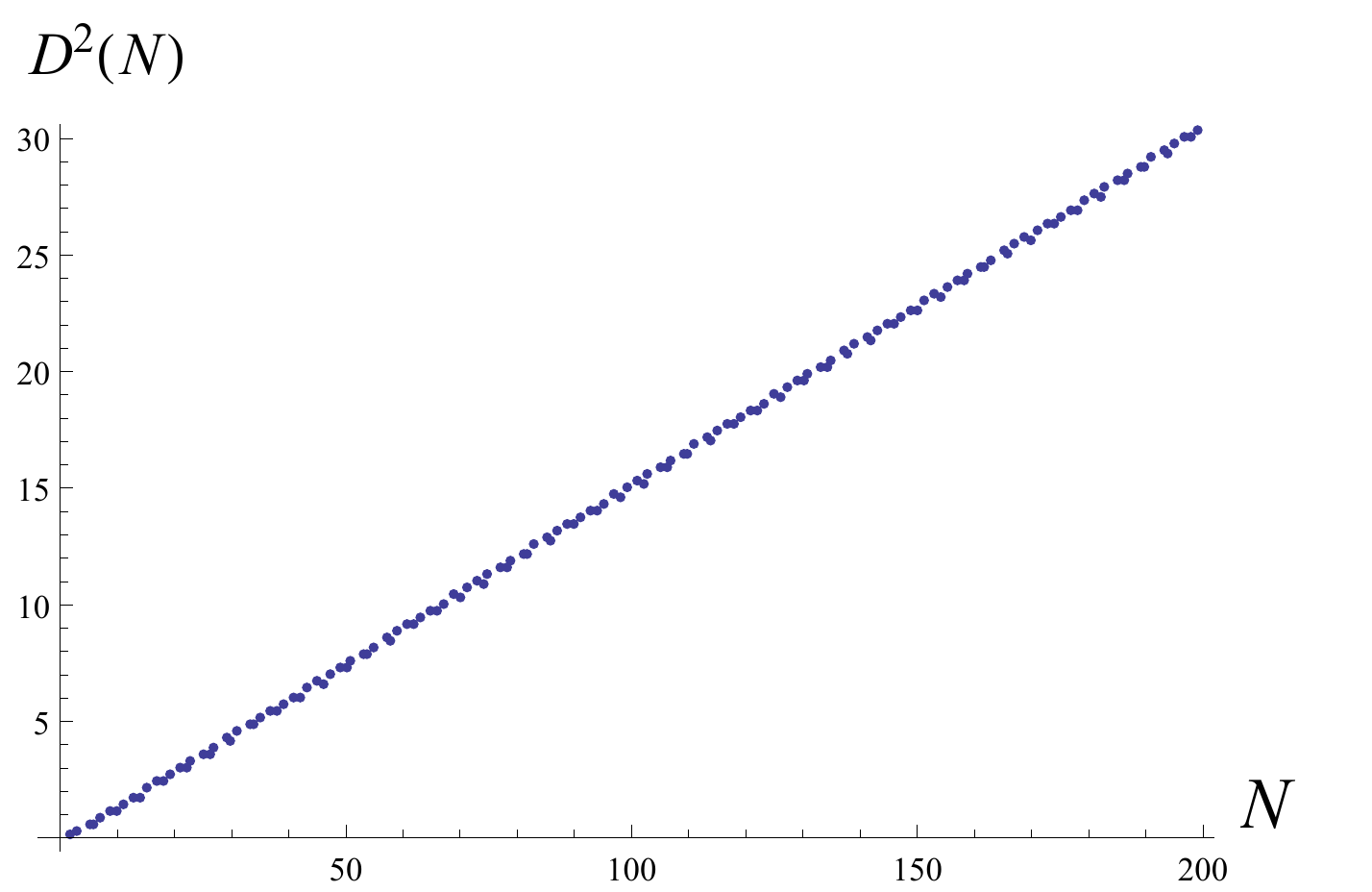}}
\\
{\color{white}0}\hspace{3cm}{ Figure 13:  Complexity v.s. lattice site number $N$}
\\
\\
It shows that the complexity increases with  site number $N$, as that in free theory. This consists with the relation :  complexity=volume (CV) conjecture \cite{Susskind1403} since in one dimension the volume is proportional to  site number $N$.  To plot figure 13 (and following figures)  we choose the scale of $\omega=1$ and use the following values : ${N_i}=1$, $\omega_f=1$, $\lambda=0.1$. The dependence of  complexity on ${N_i}$,  $\omega_f$, or $\lambda$  is illuminated  in the following figures.

(2) We plot in figure 14 the complexity for various excited state ${N_i}$. 
\\
\\
\scalebox{0.5}{\hspace{6cm}\includegraphics{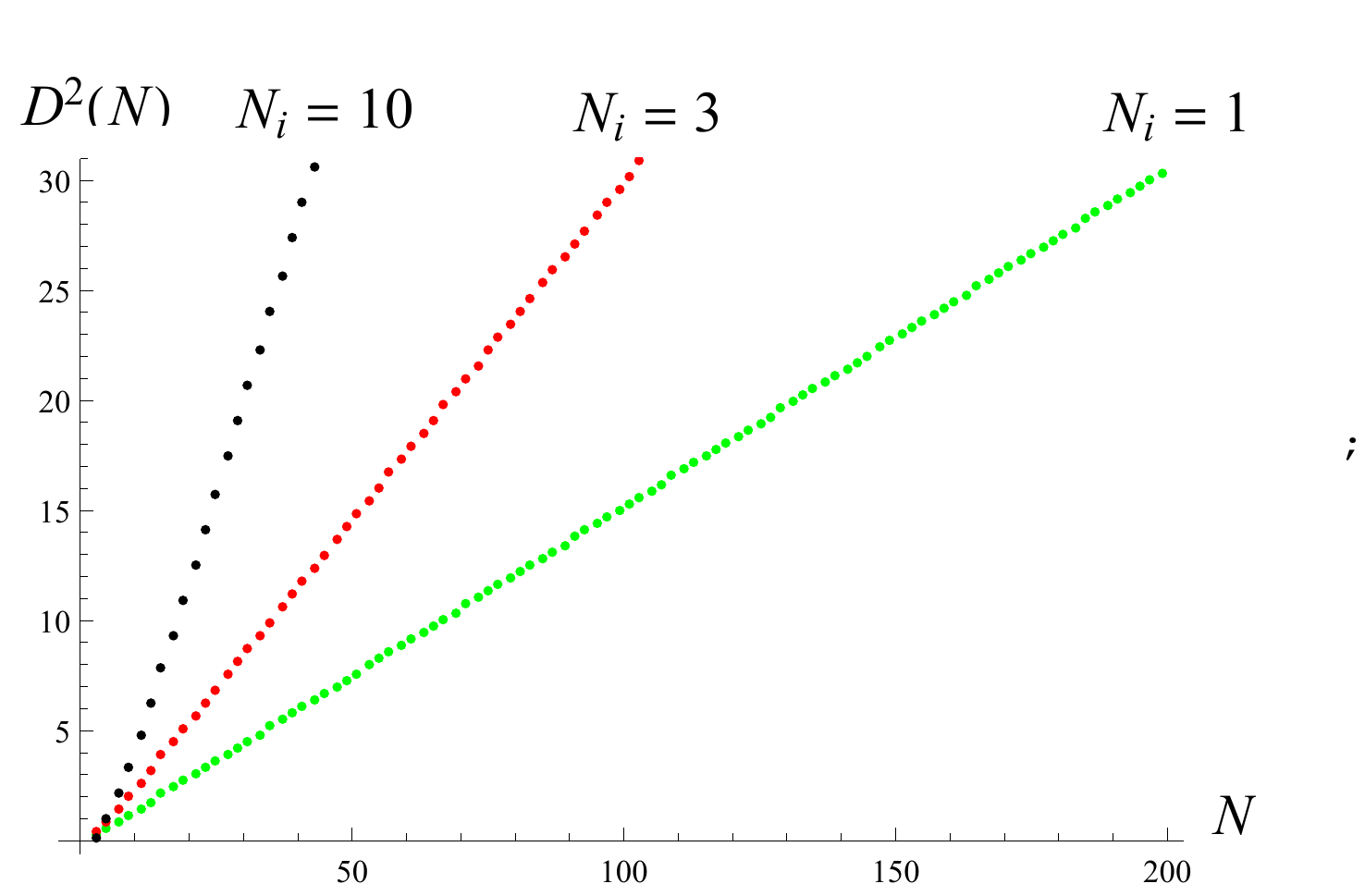}}
\\
{\color{white}0}\hspace{3cm}{ Figure 14:  Complexity v.s. excited state ${N_i}$}
\\
\\
It shows that the complexity becomes larger in higher excited state, as that in free theory.

(3) We plot in figure 15 the complexity for various  reference state frequency  $\omega _f$.
\\
\\
\scalebox{0.5}{\hspace{6cm}\includegraphics{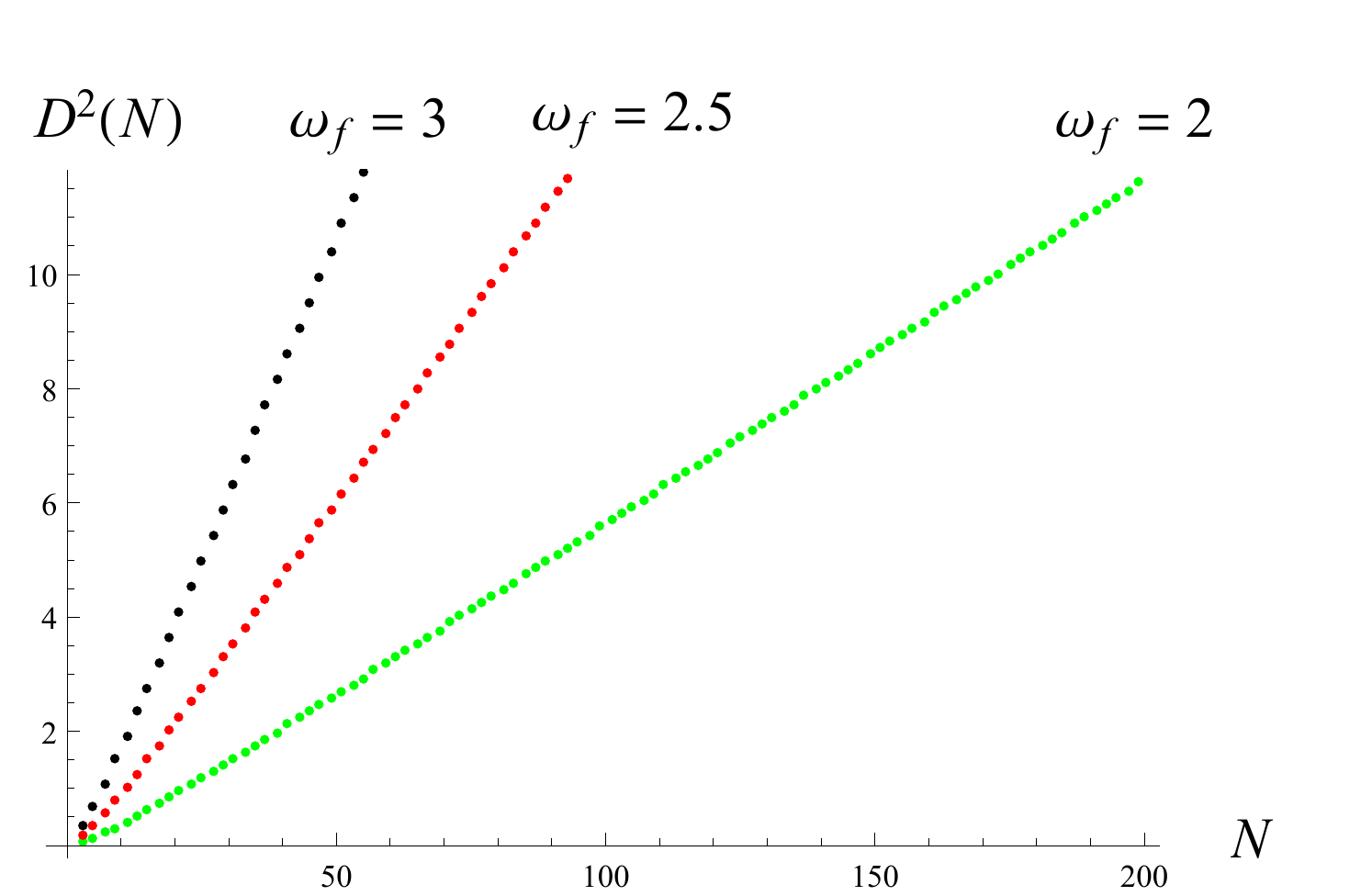}}
\\
{\color{white}0}\hspace{3cm}{ Figure 15:  Complexity v.s. reference state frequency $\omega _f$}
\\
\\
 It shows that the complexity becomes larger for large $\omega _f$, as that in free theory.

(4) We plot in figure 16 the complexity for various  coupling constant  $\lambda$ in lattice $\lambda\phi^4$ theory.
\\
\\
\scalebox{0.6}{\hspace{1cm}\includegraphics{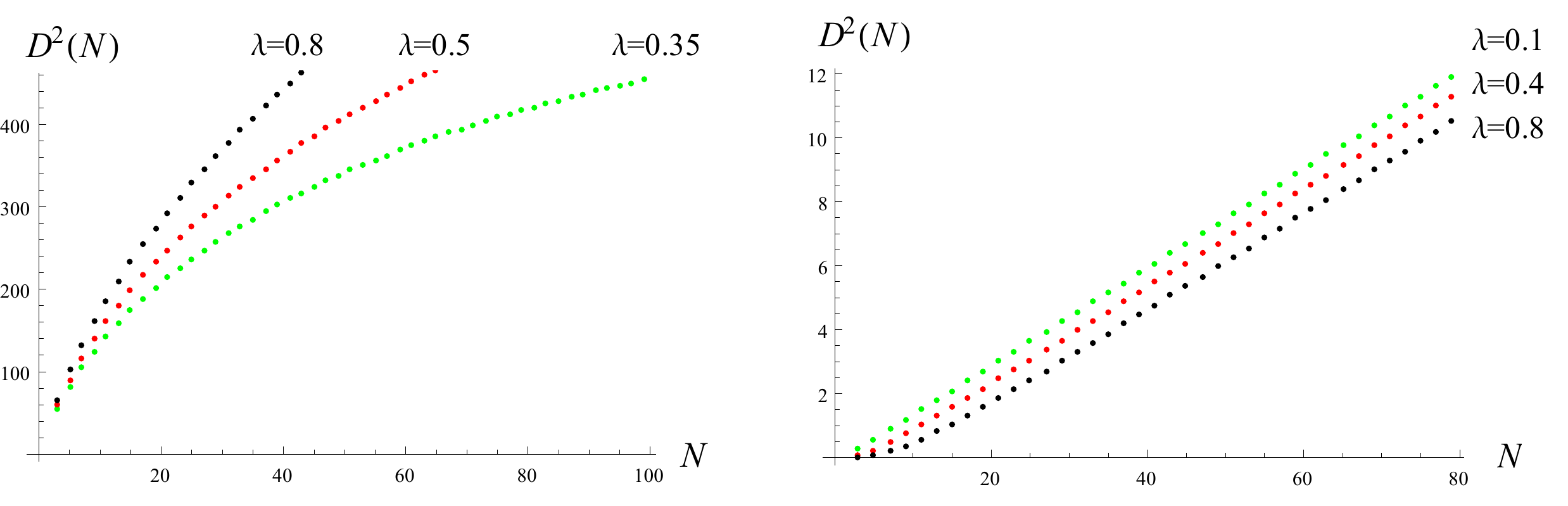}}
\\
{\hspace{0cm}{ Figure 16 :  Complexity v.s. coupling constant  $\lambda$. Left-hand diagram is that with $\omega=10,~\omega_f=0.01$. Right-hand diagram is that with $\omega=\omega_f=1$}. 
\\
\\
It shows that the complexity may increase or decrease while increasing coupling constant $\lambda$.  The property of  how the complexity depends on  $\lambda$ can be see, for example, from eq.(\ref{R2s1}) and eq.(\ref{R2s2}).  The interaction correction  to complexity in the two relations  is proportional the coefficient of ${3\lambda\over 2\omega_f}$, which is negative  for large $\omega_f$ and become positive for small $\omega_f$.  Figure 16 is consistent with this argument.
\section{Concluding Remarks}

We adopt  operator approach to compute the complexity of the lattice $\lambda\phi^4$ scalar theory.  
A perturbation algorithm has been developed for computing the complexity to obtain the  general formulae (\ref{F1}), (\ref{F2}), and (\ref{F3}) which can be used to obtain higher-order complexity of excited states for any $N$ lattice sites. The interaction correction  to complexity may be positive or negative depending on the magnitude of reference-state frequency.

We conclude the paper by  the remark : Our algorithm is based on a simple relation 
\be
\lambda a_j^\dag a_j a_j^\dag a_j\rightarrow \lambda  N_j\, a_j^\dag a_j\rightarrow  \lambda  N_jR_j^{(n-1)}\, a_j^\dag a_j
\ee
in which  the first arrow is due to the perturbation property while the second one is use to calculate the complexity. The relation   is  explained in sec.3.2. The similar relation could be found  in many other systems.   For examples  :

$\bullet$ It is easily to see that our method could be used in interacting Fermion theory.

$\bullet$ For the  theory which has two  different field operators $a_j$ and $b_j$ and associated interaction  is $\lambda \,\phi^2\,\xi^2$ the relation will become
\be
\lambda a_j^\dag a_j b_j^\dag b_j\rightarrow {\lambda\over 2}  N^{(b)}_j\, a_j^\dag a_j+{\lambda\over 2}  N^{(a)}_j\, b_j^\dag b_j\rightarrow {\lambda\over 2}  N^{(b)}_jR_j^{(b)(n-1)}\,\, a_j^\dag a_j+{\lambda\over 2}  N^{(a)}_jR_j^{(a)(n-1)}\,\, b_j^\dag b_j
\ee
in which the fields $\phi$ and $\xi$ could be Boson or Fermion field. 

$\bullet$  For the  $\lambda \phi^6$ theory the relation will become    
\be
\lambda a_j^\dag a_j a_j^\dag a_j a_j^\dag a_j\rightarrow \lambda  (N_j)^2\, a_j^\dag a_j\rightarrow  \lambda  (N_jR_j^{(n-1)})^2\, a_j^\dag a_j
\ee
Of course, the associated diagrams and basic rules in each case shall be slightly modified. 

In this way, our algorithm can be applied to many quantum field theories and several many-body models in  condense matter.  We will study the problem in the next series of paper.
\\
\begin{center} 
{\bf  \large References}
\end{center}

\end{document}